\documentclass[aps,prb,twocolumn,superscriptaddress,floatfix,showpacs]{revtex4}
\usepackage{amsmath,amssymb}
\usepackage{mathtools}
\usepackage{graphicx}
\usepackage{feynmf}
\usepackage{array}	
\usepackage{dsfont}
\usepackage{ulem}

\begin{document}

\title{Trading coherence and entropy by a quantum Maxwell demon}

\author{A.V.\ Lebedev}
\affiliation{Theoretische Physik, Wolfgang-Pauli-Strasse 27, ETH Zurich,
CH-8093 Z\"urich, Switzerland}

\author{D.\ Oehri}
\affiliation{Theoretische Physik, Wolfgang-Pauli-Strasse 27, ETH Zurich,
CH-8093 Z\"urich, Switzerland}

\author{G.B.\ Lesovik}
\affiliation{L.D.\ Landau Institute for Theoretical Physics RAS,
Akad.\ Semenova av.~1-A, Chernogolovka, 142432, Moscow Region, Russia}
\affiliation{Theoretische Physik, Wolfgang-Pauli-Strasse 27, ETH Zurich,
CH-8093 Z\"urich, Switzerland}

\author{G.\ Blatter}
\affiliation{Theoretische Physik, Wolfgang-Pauli-Strasse 27, ETH Zurich,
CH-8093 Z\"urich, Switzerland}

\date{\today}

\begin{abstract}
The Second Law of Thermodynamics states that the entropy of a closed system is
non-decreasing. Discussing the Second Law in the quantum world poses new
challenges and provides new opportunities, involving fundamental
quantum-information-theoretic questions and novel quantum-engineered devices.
In quantum mechanics, systems with an evolution described by a so-called unital
quantum channel evolve with a non-decreasing entropy.  Here, we seek the
opposite, a system described by a non-unital and, furthermore,
energy-conserving channel that describes a system whose entropy decreases with
time. We propose a setup involving a mesoscopic four-lead scatterer augmented
by a micro-environment in the form of a spin that realizes this goal. Within
this non-unital and energy-conserving quantum channel, the micro-environment
acts with two non-commuting operations on the system in an autonomous way. We
find, that the process corresponds to a partial exchange or swap between the
system and environment quantum states, with the system's entropy decreasing if
the environment's state is more pure.  This entropy-decreasing process is
naturally expressed through the action of a quantum Maxwell demon and we
propose a quantum-thermodynamic engine with four qubits that extracts work from
a single heat reservoir when provided with a reservoir of pure qubits.  The
special feature of this engine, which derives from the energy-conservation in
the non-unital quantum channel, is its separation into two cycles, a working
cycle and an entropy cycle, allowing to run this engine with no local waste
heat.
\end{abstract}


\maketitle

\section{Introduction}\label{sec:intro}

\begin{figure}[htbp]
\begin{center}
\includegraphics[width=6truecm]{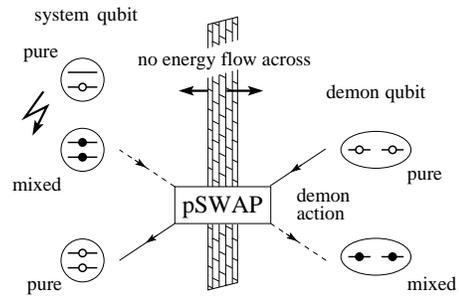}
\caption{Purification of an energetically isolated system with the help of a
quantum demon.  The system qubit on the left undergoes a transition (arrow) to a
mixed state (full circles), e.g., by decoherence or through thermal
excitation. The demon makes use of a pure demon qubit (right, open circles)
and swaps its state with the system qubit, thereby transferring entropy or
coherence but no energy.}
\label{fig:setup}
\end{center}
\end{figure}

The conversion of heat into useful energy or work is at the very heart of the
Second Law of Thermodynamics\cite{kelvin:1882}, rendering the design and
functionality of thermodynamic engines a recurrent topic.  Although dealing
with such seemingly prosaic issues as the efficiency of a machine
\cite{carnot:1824}, the Second Law implies drastic consequences, telling which
processes are allowed to occur in nature, an example being the requirement of
non-decreasing entropy in an isolated system. New challenges appear when
taking the Second Law of Thermodynamics into the quantum regime. The topic has
likewise caught the interest of the quantum-information and
quantum-engineering communities, with recent works addressing both fundamental
\cite{horodecki:2013,brandao:15,skrzypczyk:14,linden:10,brunner:2012} and
practical \cite{lloyd:97,scully:01,quan:06,quan:07,venturelli:13} issues.
Equally pertinent to the topic is Maxwell's demon
\cite{maxwell:1871,szillard:29} and its taming by Landauer's principle
\cite{landauer:61,bennett:82,bennett:03}, adding the concept of information to
the discussion \cite{deffner:13,horowitz:14}.  In this paper, we invoke a
quantum demon that makes use of quantum purity to decrease the entropy and
increase the coherence of an energetically isolated system. Starting from a
fundamental observation in quantum information theory, stating that the
entropy is non-decreasing in a unital quantum channel \cite{holevo:12}, we
search for the characteristics of a non-unital quantum channel that allows for
a maximal decrease of entropy in an isolated system. We find that such a
non-unital quantum channel involves a quantum demon that is swapping external
pure quantum states against mixed system states, see Fig.\ \ref{fig:setup}. We
present three specific examples for such demons, two mesoscopic systems using
a spin or a double quantum-dot as part of the demon that imprints a coherent
state on an incoherent electron (and thus decreases the electron's entropy),
and a four-qubit system constituting a quantum thermodynamic engine with
separated energy and entropy cycles.

A fully isolated quantum system evolves unitarily and hence its entropy
remains constant, rendering the Second Law a triviality. A non-trivial setting
is defined by an open quantum system.  Here, we have in mind a system that is
energetically isolated (and keeps a fixed particle number) but can decrease
its entropy through entanglement with its environment.  This entanglement
between the system and the environment can be induced with the help of a
phase-like interaction.  Importantly, such a phase-exchange mechanism does not
require an exchange of energy, thus modifying the concept of classical
isolation when dealing with an open quantum system. Indeed, it is well known
that typical quantum systems become more rapidly entangled through a
phase-exchange process than through a relaxation process, as prominently
expressed by the scale separation of the corresponding relaxation times for
phase ($T_2$) and energy ($T_1$) in quantum engineered systems.  Extensions of
the Second Law accounting for the presence of classical (as opposed to
quantum) correlations between the system and an information reservoir has
recently been discussed in Refs.\ [\onlinecite{deffner:13,horowitz:14}].

In a typical situation, the quantum system interacts with a large environment
that induces dephasing, eliminating the off-diagonal components of the
system's density matrix and increasing its entropy. However, if the
environment is a small and controllable quantum system, we will show below that
one can tune the interaction between the system and the environment in such a
way as to decrease the system's entropy.  Furthermore, using a phase exchange
mechanism without energy exchange with the environment, such a decrease in
entropy leads to an apparent contradiction to the traditional classical
formulation of the Second Law. In our discussion below, this entropy decrease
will be achieved by a quantum demon using a SWAP operation in order to
exchange purity (or coherence) between the micro environment and the quantum
system. Including the micro environment in the entropy balance of the grand
system reestablishes the Second Law (via Landauer's principle) in a proper
way.

The standard description of the above situation (an open quantum system
initially disentangled from its environment) is given by a {\it quantum
channel} $\Phi$. In its mathematical formulation, this is a completely
positive, trace-preserving map that transforms the initial density matrix
$\hat\rho_0$ of the quantum system to a new density matrix $\hat\rho =
\Phi(\hat\rho_0)$. In our physical context, $\Phi = \Phi_t$ describes the
evolution of the system's density matrix, $\hat\rho_t =
\mathrm{Tr}_\mathrm{env} [\hat{U}(t,t_0) \hat\rho_0 \hat{U}^\dagger(t,t_0)]
\equiv \Phi_t (\hat\rho_0)$, under the inclusion of the environment
($\hat{U}(t,t_0)$ denotes the evolution of the grand system). The entropy of
such an open quantum system is not conserved, $S[\Phi(\hat\rho_0)] \neq
S(\hat\rho_0)$, and actually can both increase and decrease. Quantum
information theory then offers the remarkable statement that the entropy gain
is bounded from below, $S[\Phi(\hat\rho_0)] - S(\hat\rho_0) \geq -\mathrm{Tr}
\bigl\{[\Phi(\hat\rho_0)] \,\ln \Phi(\mathds{1}))\bigr\}$.  It follows that a
special subclass of quantum channels which preserve the identity operator,
$\Phi(\mathds{1}) = \mathds{1}$, has a non-negative entropy gain,
$\ln(\mathds{1}) = 0$ and hence $S[\Phi(\hat\rho_0)] - S(\hat\rho_0) \geq 0$.

A quantum system whose evolution can be described by a an identity-preserving
quantum channel evolves with a non-decreasing entropy, $S[\Phi(\hat\rho_0)] \geq
S(\hat\rho_0)$; such a quantum channel is called unital.  It turns out that
for finite dimensional quantum systems, unitality is not only a sufficient but
also a necessary condition for a non-negative entropy gain.  Indeed, if a
$N$-dimensional quantum system evolves with non-decreasing entropy for any
initial state $\hat\rho_0$, then it does so also for the completely chaotic
state $\hat\rho_c = \mathds{1}/N$ and therefore $S[\Phi(\hat\rho_c)] \geq
S(\hat\rho_c)$. Since the chaotic state $\hat\rho_c$ has maximal possible
entropy $k_\mathrm{\scriptscriptstyle B} \ln(N)$ one has $S[\Phi(\hat\rho_c)]
= k_\mathrm{\scriptscriptstyle B} \ln(N)$ and hence $\Phi(\hat\rho_c) =
\mathds{1}/N$, that proves the unitality of $\Phi$. These considerations tell
us that in order to find an isolated finite-dimensional quantum system whose
entropy decreases under its evolution, we have to search for a non-unital
energy-conserving quantum channel and a suitable initial state. Note, that
unitality does not imply energy conservation of a quantum channel. Indeed,
consider a quantum channel $\Phi(\hat\rho) =
\hat{P}\bigl(\mbox{diag}(\hat\rho)\bigr)$ which eliminates all off-diagonal
elements of the density matrix and imposes an arbitrary permutation $\hat{P}$
of its diagonal elements. This quantum channel is unital, however, the
permutation of the diagonal elements in an energy representation does not
conserve the total energy of the quantum system.

The existence of non-unital energy-conserving quantum channels that reduce the
entropy has been pointed out in Ref.\ [\onlinecite{sadovskyy:14}]: scattering an
electron in a three-terminal mesoscopic conductor suitably interacting with a
spin, reduces the entropy of the outgoing state with respect to the incoming
one.  Below, we investigate the functionality of such a quantum channel in
more detail and find conditions and specific examples that lead to a maximal
decrease of entropy.  We will present two mesoscopic scattering setups that
define such energy-conserving non-unital quantum channels with initial states
that evolve with maximal decreasing entropy. In both examples, we consider a
four-lead reflectionless beam splitter and a general incoming electronic state
$\hat\rho_\mathrm{in}$ that is scattered by the symmetric splitter. The system
is augmented by a qubit acting as the microscopic environment. The interaction
between the qubit and the scattering electron is arranged in such a way as to
define a non-unital quantum channel that decreases the entropy of the outgoing
state $\hat\rho_\mathrm{out}$ with respect to the incoming one,
$S(\hat\rho_\mathrm{out}) \leq S(\hat\rho_\mathrm{in})$, without exchange of
energy between the electron and the qubit. In the first example, the qubit is
given by a spin, prepared in a suitable state, that interacts with electrons
propagating in two of the four leads, thereby executing two non-commuting
operations on the corresponding components---the non-commutativity of the two
operations as derived from the non-unitality condition is a central feature of
this setup. In the second example, we replace the spin by a double-quantum-dot
that is more easily manipulated in a realistic system. Here, the quantum dot
exerts the identical operation on the electron, once in the incoming and a
second time in the outgoing lead, however, suitably rotating the double-dot's
state in between the two interaction events renders the two operations on the
electron effectively non-commuting. Within the mesoscopic transport setting,
the reduced entropy in the outgoing lead corresponds to an increase in
coherence of the outgoing electron state, where the attained coherence is
provided by the demon.  The concept may be useful in locally generating a
coherent state of a flying qubit using the purity of a (usually more stable)
stationary qubit.  In order to test the functionality of such a device, we
propose to analyze this imposed coherence in a Mach-Zehnder setup.

Analyzing the functionality of these devices in terms of a quantum algorithm,
we find that the various steps in the protocol exchange the quantum states of
the system and the demon. Rather than a standard SWAP
\cite{lloyd:97,quan:06,quan:07}, our considerations lead to a partial SWAP
operation (pSWAP) that exchanges states in one sector of the Hilbert space,
while producing an incomplete SWAP (a SWAP up to a NOT operation) in the
remaining part of the Hilbert space. While a conventional SWAP operation also
achieves the functionality of the non-unital quantum channel, its physical
implementation is more demanding within the context of the present paper.

The purification of a quantum state in a demon-assisted process can be used in
the construction of a quantum thermodynamic engine.  The discussion of quantum
thermodynamic processes has continuously progressed over the past two decades,
see [\onlinecite{mahler:14}] for a recent review.  Quantum thermodynamic
machines usually involve hot and cold (switchable) thermal reservoirs in
combination with a (spectrally tunable) quantum system as an operating medium,
see, e.g., Refs.\ [\onlinecite{scovil:59,palao:01,feldmann:03,segal:06,pekola:16}],
often operating in an autonomous manner \cite{tonner:05}. Here, we focus on a
class of machines that make use of a SWAP operation between states of
different purity. This operation is conveniently described within the
framework of Maxwell's demon \cite{maxwell:1871,szillard:29}. The demon's
original\cite{maxwell:1871,szillard:29} resource is the ability to distinguish
particle motion such as to transfer them unidirectionally between two adjacent
volumes, thereby reducing the system's entropy without investing work.  The
violation of the Second Law in the restricted system then is cured by going
over to a larger system including the Maxwell demon.  This is the essence of
Landauer's principle, \cite{landauer:61,bennett:82,bennett:03} stating that
resetting the demon's memory costs an entropy $k_{\rm\scriptscriptstyle B} \ln
2$ per decision. This entropy has to be picked up by a second reservoir, e.g.,
at a lower temperature $T_0$, and the putative perpetuum mobile of the second
kind transforms into a conventional Carnot machine.

Through swapping the (less pure) system- with a purer demon state, our quantum
demon provides the entropy required to have the machine run with only one
thermal reservoir, thus replacing Maxwell's classical demon in a quantum
thermodynamic engine.  This idea has been introduced early on by
Lloyd\cite{lloyd:97} in his construction of a spin-based NMR demon and a
proposal for a quantum thermodynamic engine.  An atom-based quantum heat
engine transforming heat from one reservoir into work has been proposed by
Scully \cite{scully:01}, with the negentropy \cite{scully:01} consumed in the
cycle corresponding to a reservoir of demon qubits within our language.
Quantum heat engines assisted by a similar quantum demon have been described
by Quan {\it et al.} \cite{quan:06}, see also Ref.\ [\onlinecite{quan:07}],
and more recently by Diaz de la Cruz and Martin-Delgado \cite{delacruz:14}.
The SWAP operation is a central element in algorithmic cooling, e.g., of spins
in NMR spectroscopy or for NMR based quantum computation
\cite{boykin:02,fernandez:04,brassard:14} as well as in the discussion of the
smallness of quantum thermodynamic machines \cite{linden:10}.

The SWAP-based quantum heat engines proposed so far involve different
characteristics and functionalities. E.g., the engines in Refs.\
[\onlinecite{lloyd:97,quan:06,quan:07}] involve working- and demon qubits with
gapped spectra and a SWAP between two thermal states. In such a setup the
injected heat is only partly transformed into work. This is due to the energy
exchange between working- and demon qubits in the SWAP operation, always
reducing the work (or incoming heat) by a waste heat. A setup which does not
involve an energy exchange with the demon qubit has been proposed in Ref.\
[\onlinecite{delacruz:14}]. This engine makes use of energy-degenerate
working- and demon qubits, with the work extracted in a
polarization/depolarization process, which requires a slow (adjabatic)
process.

In our version of a demon-assisted thermodynamic engine, we make use of a
partial SWAP (pSWAP) operation exchanging the pure state of an
energy-degenerate demon qubit (the dit) with the (non-degenerate) working
qubit's (the wit's) thermally excited state in an autonomous process.  The
exchange of quantum states in the pSWAP operation transforms the wit's thermal
energy into the directed energy of a pure excited wit state. The wit's energy
then can be extracted and used with the help of suitable quantum
manipulations, while the wit's entropy is carried away by the dit after the
pSWAP.  While the pSWAP exchanges the entropies (or purities) of the wit and dit,
the wit's and dit's energies are separately conserved.  The separation into
distinct energy and entropy cycles with no energy transfer in between then is
the most interesting feature of our quantum-thermodynamic engine.
In particular, considering only the energy cycle of the wit, it turns out that
the local heat-to-work transfer operates with unit efficiency, i.e., the heat
absorbed from the thermal reservoir by the wit can be fully converted into
directed work. Furthermore, the entropy cycle where the dit is reprepared for
the next round of operation can be run separately, in time or space, from the
working cycle of the engine; alternatively, the dits can be provided from a
previously prepared reservoir of dit states. On the other hand, combining the
outcome of both the energy and entropy cycles, i.e., accounting for the state
changes of both the wit and the dit, the Second Law and Landauer's Principle
are fully respected and the resulting overall efficiency is smaller than the
one of a Carnot machine.

In the following section \ref{sec:spin}, we introduce an electronic scattering
problem which includes the interaction with an auxiliary spin (our first type
of demon) and formulate the process in the language of a quantum channel to
derive the conditions for an energy-preserving and entropy-decreasing
non-unital quantum channel.  In section \ref{sec:dd}, we proceed with the discussion
of a more practical setup using a quantum double-dot as the demon qubit and
proceed to describe the functionality of this non-unital quantum channel in terms of a
quantum circuit, see Sec.\ \ref{sec:qcd}. Section \ref{sec:qte} is
devoted to our quantum thermodynamic engine with its special property of
separate energy and entropy cycles. In section \ref{sec:sc}, we summarize our
findings and conclude.

\section{Entropy reduction in a non-unital quantum channel}\label{sec:spin}

\begin{figure}[htbp]
\begin{center}
\includegraphics[width = 8truecm]{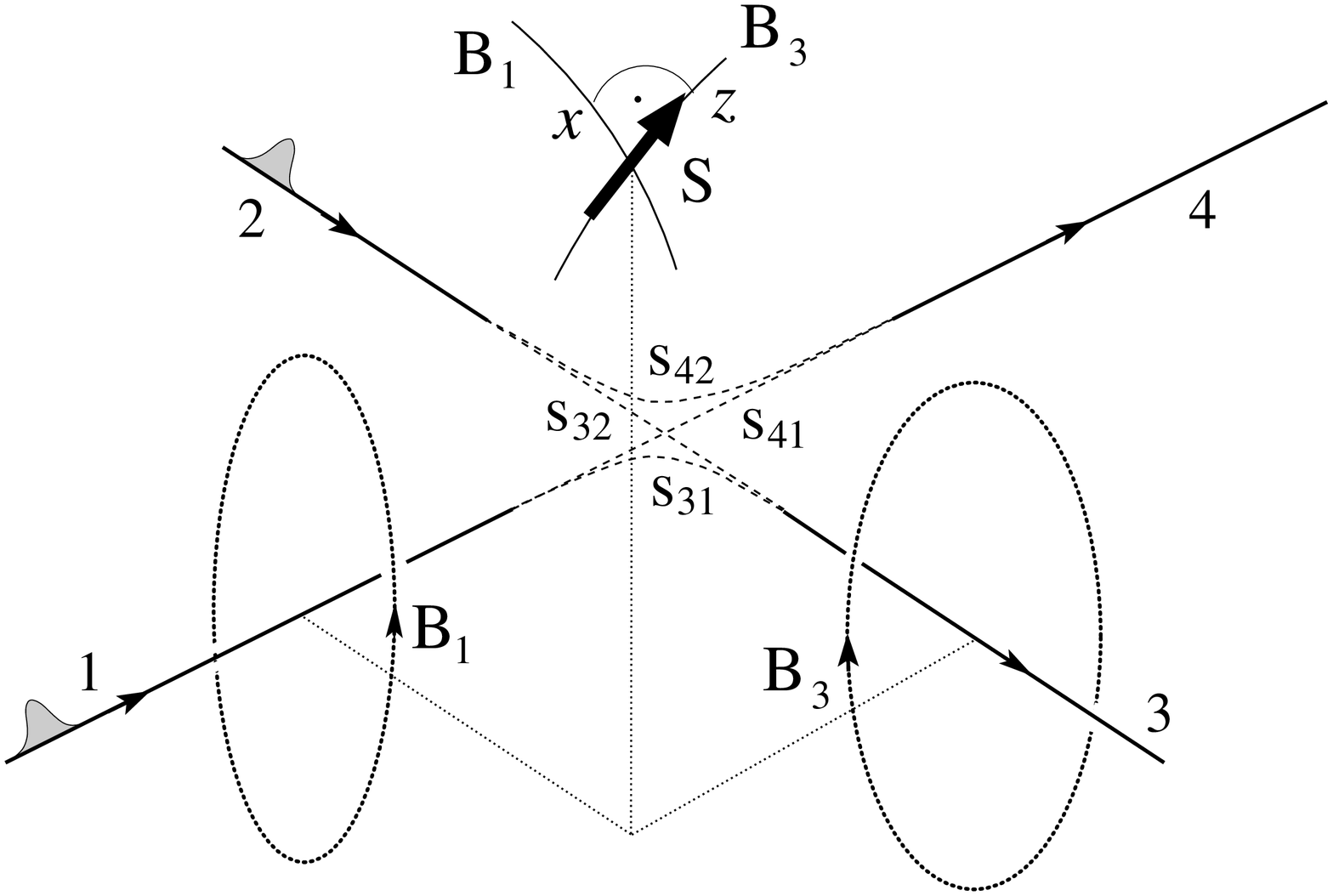}
\caption{Non-unital quantum channel realizing maximal purification by two
consecutive orthogonal rotations $\hat{u}_1 = \sigma_x$ and $\hat{u}_3=
\sigma_z$.  The 3D sketch illustrates the scattering of an electron incident
from leads 1 and 2 and scattered into the leads 3 and 4; $s_{ij}$ denote the
scattering amplitudes. The micro-environment is defined through the spin ${\bf
S}$ which interacts with the magnetic fields ${\bf B}$ generated by the
electron travelling in the close-by leads 1 (generating the field ${\bf B}_1$)
and 3 (generating the field ${\bf B}_3$); we ignore the interaction with
currents in the distant leads 2 and 4.  The geometry of the leads and the spin
position is chosen such that the magnetic fields of the currents in leads 1
and 3 are orthogonal. The initial (or operational) spin states
$|\!\!\uparrow\rangle$ and $|\!\!\downarrow\rangle$ are chosen parallel to $z$
(the field generated by a current in lead 3) such that the passage of the
electron in lead one flips the spin and generates the unitary $\hat{u}_1$. The
passage of the electron through the lead 3 induces a rotation around $z$ and
adds the phases as required for the unitary $\hat{u}_3$.}
\label{fig:spin-setup}
\end{center}
\end{figure}

We consider a single electron, our system, propagating through a
reflectionless beam splitter and interacting with a localized quantum spin
assuming the role of the micro environment, see Fig.\ (\ref{fig:spin-setup})
(in more general terms, this can be viewed as a flying system-qubit
interacting with a environment-qubit).  Our goal is to find a simple
realization of a non-unital quantum channel that decreases the system's
entropy when propagating from the input to the output leads. The interaction
between the electron and the spin is mediated through the magnetic field
generated by the electron's motion. An electron propagating through the lead
$\alpha$ induces a unitary rotation $\hat{u}_\alpha$ of the spin.  The
electron scattering is described by the unitary scattering matrix $\hat{s}$,
$|\beta\rangle = \sum_\alpha s_{\beta\alpha} |\alpha\rangle$, where
$|\alpha\rangle$ describes the localized electron wave-packet moving in the
incoming ($\alpha = 1,2$) or outgoing ($\beta = 3,4$) leads of the beam
splitter. The incoming leads are defined by the electron's incoming state and
are clearly distinguished from the outgoing state through the reflectionless
property of the beam splitter.

We assume an initial state of the grand system `electron plus spin' in a
product form $\hat{R} = \hat\rho\, \otimes\, \hat{r}$, with the initial density
matrices of the electron $\hat\rho = \sum_{\alpha \alpha^\prime}
\rho_{\alpha\alpha^\prime} \,|\alpha\rangle \langle \alpha^\prime|$ and the
spin $\hat{r}$ to be determined. After the scattering, the density matrix of
the grand system has the form
\begin{equation}
   \mathcal{D}\hat{R} = \!\!\!\! \sum_{\alpha\alpha^\prime\beta\beta^\prime} \!\!
   s_{\beta\alpha} \rho_{\alpha\alpha^\prime} s_{\beta^\prime\alpha^\prime}^*
   \, |\beta\rangle \langle \beta^\prime| \otimes \bigl[ \hat{u}_\beta\hat{u}_\alpha
   \,\hat{r}\, \hat{u}_{\alpha^\prime}^\dagger \hat{u}_{\beta^\prime}^\dagger \bigr]
\end{equation}
and the resulting density matrix of the electron, $\mathcal{D} \hat\rho =
\mbox{Tr}_\mathrm{env}\{ \mathcal{D}\hat{R}\}$, is given by
\begin{equation}
  [\mathcal{D}\rho]_{\beta\beta^\prime} = \sum_{\alpha\alpha^\prime}
   s_{\beta\alpha} \rho_{\alpha\alpha^\prime} s_{\beta^\prime\alpha^\prime}^*
   \, \mbox{Tr}_\mathrm{env}\{ \hat{r}\, \hat{u}_{\alpha^\prime}^\dagger
   \hat{u}_{\beta^\prime}^\dagger \hat{u}_\beta \hat{u}_\alpha\}.
   \label{eq:channel}
\end{equation}
The transformation $\mathcal{D}$ of the electron density matrix defined in
Eq.\ (\ref{eq:channel}) defines a quantum channel $\Phi$. In the absence of an
external magnetic field, the electron and spin do not exchange energy, i.e.,
the electron represents a thermodynamically isolated system in the classical
sense. In order to discuss the unitality of this quantum channel, we analyse
the evolution of the electron's chaotic quantum state $[\hat\rho_c]_{\alpha
\alpha'} =\delta_{\alpha\alpha'}/2$ (or short $\hat\rho_c = \mathds{1}/2$)
which assumes the form
\begin{equation}
   \Phi(\rho_c) = \mathcal{D} \rho_c = \frac12
    \left( \begin{array}{cc}
    1&\gamma\\ \gamma^*&1
    \end{array}\right),
\end{equation}
where $\gamma = s_{31}s_{41}^* \mbox{Tr}\bigl\{ \hat{r}
\bigr[\hat{u}_1^\dagger \hat{u}_4^\dagger \hat{u}_3 \hat{u}_1 -
\hat{u}_2^\dagger \hat{u}_4^\dagger \hat{u}_3 \hat{u}_2\bigr]\bigr\}$ and the
electron's entropy gain is given by $\Delta S(\hat\rho) =
S(\mathcal{D}\hat\rho) - S(\hat\rho) = - [(1+|\gamma|) \ln(1+|\gamma|) +
(1-|\gamma|) \ln(1-|\gamma|)]/2$. The entropy gain is negative and assumes its
absolute maximal value $\ln 2$ at $|\gamma| = 1$.  This value indeed can be
realized for a specific choice of the device parameters. First, the maximum
value of $|s_{31}s_{41}^*|$ is 1/2 and requires a symmetric beam splitter.  Next,
we assume that the electron interacts with the spin only if it propagates in
the bottom leads $1$ and $3$. Then, $|\gamma| = \bigl| \mbox{Tr}\bigl\{
\hat{r} \bigl[\hat{u}_1^\dagger \hat{u}_3 \hat{u}_1 - \hat{u}_3\bigr]\bigr\}/2
\bigr|$ and its non-vanishing requires non-commuting spin rotations
$\hat{u}_1$ and $\hat{u}_3$. The maximal value of $|\gamma| = 1$ is attained
if there is a pure spin state $\hat{r} = |\!\!\uparrow\rangle \langle
\uparrow\!\!|$ for which the unitary operations $\hat{u}_1$ and $\hat{u}_3$
satisfy the two conditions $\langle \uparrow\!\!| \hat{u}_1^\dagger \hat{u}_3
\hat{u}_1|\!\!\uparrow\rangle = e^{i\phi}$ and $\langle \uparrow\!\!|
\hat{u}_3 |\!\!\uparrow\rangle = - e^{i\phi}$ with an arbitrary phase $\phi$.
The second condition requires that $\hat{u}_3|\!\!\uparrow\rangle =
-e^{i\phi}|\!\!\uparrow\rangle$ (and $\hat{u}_3 |\!\!\downarrow\rangle =
e^{i\phi}|\!\!\downarrow\rangle$ when choosing $\hat{r} = |\!\!\downarrow
\rangle \langle \downarrow\!\!|$, see below). In analyzing the first
condition, we start with a general ansatz for $\hat{u}_1$, $\hat{u}_1|\!\!
\uparrow\rangle = a |\!\!\uparrow\rangle + b|\!\!\downarrow\rangle$ with
$|a|^2 + |b|^2 = 1$.  The first condition requires that $\langle \uparrow\!\!|
\hat{u}_1^\dagger \hat{u}_3 \hat{u}_1|\!\!\uparrow\rangle = -|a|^2 e^{i\phi} +
|b|^2 e^{i\varphi} = e^{i\phi}$ and hence $a=0$, $e^{i(\phi - \varphi)} = 1$.
We thus arrive at the following constraints for the unitary operations
$\hat{u}_1$ and $\hat{u}_3$ that maximize $|\gamma|$,
\begin{eqnarray}
      &&\hat{u}_1|\!\!\uparrow\rangle = e^{i\alpha}|\!\!\downarrow\rangle,
      \quad\>\>\>
      \hat{u}_1|\!\!\downarrow\rangle = e^{i\beta} |\!\!\uparrow\rangle,
      \label{eq:untr1}
      \\
      &&\hat{u}_3|\!\!\uparrow\rangle = -e^{i\phi}|\!\!\uparrow\rangle, \quad
      \hat{u}_3|\!\!\downarrow\rangle = e^{i\phi}|\!\!\downarrow\rangle,
      \label{eq:untr2}
\end{eqnarray}
where $|\!\!\uparrow\rangle$ and $|\!\!\downarrow\rangle$ are some orthogonal
spin states and $\phi$, $\alpha$, and $\beta$ are arbitrary parameters. The
conditions \eqref{eq:untr1} and \eqref{eq:untr2} mutually define the
operational states $|\!\!\uparrow\rangle$ and $|\!\!\downarrow\rangle$ and one
easily checks that $\gamma = 2 s_{31} s_{41}^* e^{i\phi}$. A similar outcome
with maximal $|\gamma|=1$ but reversed sign $\gamma = - 2 s_{31} s_{41}^*
e^{i\phi}$ appears when the spin is prepared in the orthogonal pure state
$\hat{r} = |\!\!\downarrow\rangle \langle \downarrow\!\!|$. As a consequence,
although the matrix element $\langle \downarrow\!\!| \hat{u}_1^\dagger
\hat{u}_3 \hat{u}_1 - \hat{u}_3|\!\!\uparrow\rangle = 0$ vanishes, a general
superposition state $a|\!\!\uparrow\rangle + b|\!\!\downarrow\rangle$ of the
qubit {\it does not} satisfy the condition $|\gamma| = 1$; e.g., a balanced
superposition with $|a|^2 = |b|^2 = 1/2$ will reproduce the chaotic state with
$\gamma = 0$. A possible geometry that implements this quantum channel is
sketched in Fig.\ \ref{fig:spin-setup}.

Hence, the above autonomous interaction of the electron residing in the fully
chaotic state $\hat\rho_c$ and the spin prepared in either of the pure states
$|\!\!\uparrow\rangle$ or $|\!\!\downarrow \rangle$ leads to a final electron
state with vanishing entropy, i.e., the state of the system (the electron) is
purified by the spin.  Furthermore, with the electron appearing in a pure
state after the transformation, the overall `electron plus spin' state
factorizes. Since the entropy of the grand system is conserved, the initial
electron's entropy is absorbed by the spin, with the latter appearing in the
fully chaotic state after the interaction with the electron.

Next, we generalize the observation that our spin-augmented scattering process
exchanges a fully chaotic system state with the spin's pure states $|\!\!
\uparrow \rangle$ or $|\!\!\downarrow \rangle$ to the cases of arbitrary pure
and mixed initial system states as well as mixtures of $|\!\!\uparrow\rangle$
and $|\!\!\downarrow \rangle$ spin-states---note that this list is exhaustive,
i.e., superpositions of $|\!\!\uparrow\rangle$ and $|\!\!\downarrow \rangle$
spin-states or even mixtures of such superpositions are not allowed. This
analysis will demonstrate that our spin-augmented scattering setup acts as an
autonomous quantum Maxwell demon with specific capabilities of exchanging
states between the system and the micro-environment.

First, consider the transformation of a pure initial electron state. It is
convenient to introduce the notation $|1\rangle = |\!\!\Uparrow\rangle$ and
$|2\rangle = |\!\!\Downarrow\rangle$ for the incoming states. Similarly, we
use the notation $|3\rangle = |\!\Uparrow\rangle$ and $|4\rangle =
|\!\Downarrow\rangle$ for the outgoing leads in our reflectionless scatterer
and keep in mind that we switch the meaning of the $|\!\Uparrow\rangle$ and
$|\!\Downarrow\rangle$ states when going from incoming to outgoing leads. We
choose a symmetric beam splitter in the most general form,
\begin{equation}
      \hat{s}(\theta,\eta) = 
      \left(\begin{array}{cc}
      s_{31}&s_{32} \\ s_{41}&s_{42}
      \end{array} \right)
      = \frac1{\sqrt{2}}\left(
      \begin{array}{cc}
      e^{i\theta}&-e^{-i\eta}\\e^{i\eta}&e^{-i\theta}
      \end{array} \right).
      \label{eq:scmat}
\end{equation}
Then, an arbitrary incoming state $|\phi_0\rangle = a |\!\!\Uparrow\rangle + b
|\!\!\Downarrow\rangle$ of the electron experiences the transformation
\begin{eqnarray}\label{eq:swap1}
    &&|\phi_0\rangle \!\otimes |\!\!\uparrow\rangle \to \phantom{+} 
   |\!\!\Uparrow_{xy}\rangle \!\otimes \bigl[ a e^{i\alpha}
   |\!\!\downarrow\rangle + b e^{-i(\eta+\theta)}|\!\!\uparrow\rangle\bigr],
      \\
   &&|\phi_0\rangle \!\otimes |\!\!\downarrow\rangle \to - 
   |\!\!\Downarrow_{xy}\rangle \!\otimes \bigl[ b e^{-i(\eta+\theta)}
   |\!\!\downarrow\rangle + a e^{i\beta}|\!\!\uparrow\rangle \bigr],
   \nonumber
\end{eqnarray}
where
\begin{align}\label{eq:Upxy}
   |\!\!\Uparrow_{xy}\rangle 
   &= \frac{e^{i(\phi+\theta)} |\!\!\Uparrow\rangle 
   + e^{i\eta}|\!\!\Downarrow\rangle}{\sqrt{2}},\\ 
   \nonumber
   |\!\!\Downarrow_{xy}\rangle 
   &= \frac{e^{i(\phi+\theta)} |\!\!\Uparrow\rangle 
   - e^{i\eta}|\!\!\Downarrow\rangle}{\sqrt{2}},
\end{align}
denote pure system states polarized within the equatorial $xy$-plane of the
pseudo-spin states $|\!\!\Uparrow\rangle$ and $|\!\!\Downarrow\rangle$ and
the final state of the spin is uniquely defined by the initial state
$|\phi_0\rangle$ of the electron/system.

In both cases, chaotic or pure initial system states, the final state of the
grand system (electron plus qubit) is again a product state, provided that the
qubit was initially prepared in one of the specific operating states
$|\!\!\uparrow \rangle$ or $|\!\!\downarrow\rangle$ defined the by
electron--qubit interaction, see Eqs.\ (\ref{eq:untr1}) and (\ref{eq:untr2}).
Moreover, the final states $|\!\!\Uparrow_{xy}\rangle$ and
$|\!\!\Downarrow_{xy}\rangle$ of the electron do not depend on its initial
state $|\phi_0\rangle$ and are determined only by the initial state of the
qubit, by the parameters $(\theta, \eta)$ of the beam splitter, and by the
interaction phase $\phi$. The additional phases $\alpha$ and $\beta$ appearing
in the definition of $\hat{u}_1$, Eq.\ \eqref{eq:untr1}, show up only in the
final qubit state.

In a more general situation, the initial states of both the system as well as
the spin qubit can be mixed. We start with the qubit in an incoherent mixture
of the $|\!\!\uparrow \rangle$ and $|\!\!\downarrow\rangle$ states, $\hat{r} =
p_+ {|\!\!\uparrow\rangle} \langle \uparrow\!\!| + p_-|\!\!\downarrow\rangle
\langle \downarrow\!\!|$ and the electron in a mixed state $\hat\rho$ written
in the diagonal representation $\hat\rho = p_1 |\phi_1\rangle \langle \phi_1|
+ p_2 |\phi_2\rangle \langle \phi_2|$ with $|\phi_j\rangle = a_j
|\!\!\Uparrow\rangle + b_j |\!\!\Downarrow\rangle$ and $\langle \phi_1|
\phi_2\rangle = 0$.  Then, according to Eq.\ (\ref{eq:swap1}), the initial
product state $\hat{R} = \hat\rho \otimes \hat{r}$ transforms into the sum
of product states
\begin{equation} \label{eq:density1}
  \mathcal{D} \hat{R}
  = p_+ |\!\!\Uparrow_{xy}\rangle \langle \Uparrow_{xy} 
   \!\!|\otimes \hat{r}_+  + p_- |\!\!\Downarrow_{xy}\rangle 
   \langle \Downarrow_{xy}\!\!| \otimes \hat{r}_-,
\end{equation}
where the final density matrices of the qubit $\hat{r}_\pm= p_1|\psi_{\pm
1}\rangle \langle \psi_{\pm 1}| + p_2 |\psi_{\pm 2}\rangle \langle \psi_{\pm
2}|$ are defined by the initial states of the electron, $|\psi_{j}\rangle =
b_j e^{-i(\eta +\theta)} {|\!\!\uparrow\rangle} + a_j e^{i
\alpha}|\!\!\downarrow\rangle$, $|\psi_{-j}\rangle = a_j e^{i\beta}
|\!\!\uparrow\rangle +b_j e^{-i(\eta+\theta)} |\!\!\downarrow\rangle$.

Comparing initial and final states in \eqref{eq:density1}, one notes that the
electron and spin have swapped their initial entropies.  The most extreme case
is given by an electron incoming in the chaotic state $\hat\rho = \mathds{1}/2
= ({|\!\!\Uparrow\rangle} {\langle \Uparrow\!\!|} +
|\!\!\Downarrow\rangle \langle \Downarrow\!\!| \bigr)/2$ with maximal entropy
$k_{\rm \scriptscriptstyle B} \ln 2$ and the qubit in one of the pure states
$|\!\!\uparrow\rangle$ or $|\!\!\downarrow\rangle$. Then, the product state of
the grand system, e.g., $\hat{R} = \mathds{1}/2 \otimes |\!\!\uparrow\rangle
\langle \uparrow\!\!|$, is transformed into the new product state
$\mathcal{D}\hat{R} = |\!\!\Uparrow_{xy} \rangle \langle \Uparrow_{xy}\!\!|
\otimes \mathds{1}/2$ with the electron residing in the pure state $|\!\!
\Uparrow_{xy}\rangle$.  Hence, our qubit-assisted scattering setup acts as an
autonomous quantum Maxwell demon, with the electron's original entropy
$k_{\rm\scriptscriptstyle B} \ln 2$ completely transferred to the spin qubit,
as the latter ends up in the chaotic state. We thus call the auxiliary
spin-qubit our `purifying' or `demon' qubit.  Note that this transfer of
entropy does not involve any transfer of energy, nor is there an external
supply of energy during the process.

However, the above exchange of entropy holds true only if the demon qubit has
been initially prepared in either of the pure states $|\!\!\uparrow\rangle$ or
$|\!\!\downarrow\rangle$ or an incoherent mixture thereof. For such a specific
preparation of the demon, the entropy of the system is reduced if the initial
state of the demon is more pure than the one of the system.  In the more
general situation where the initial state of the demon is a superposition
state of $|\!\!\uparrow\rangle$ and $|\!\!\downarrow\rangle$ or an arbitrary
mixed state, the above scheme does not lead to an entropy exchange between the
two subsystems. Thus, although our process does not require knowledge about
the initial system state, it does require a proper preparation of the demon
qubit.

\section{Imposed coherence on a flying qubit}\label{sec:dd}

We now study a scattering electron within an alternative setup where the spin
defining the micro-environment is replaced by a quantum double-dot. This setup
appears simpler to realize, with the action of the micro-environment involving
equal operations before and after the scattering event. The price to pay then
is an additional basis change (or rotation) of the double-dot qubit in between
the two interaction events.

\begin{figure}[h]
\begin{center}
\includegraphics[width=7truecm]{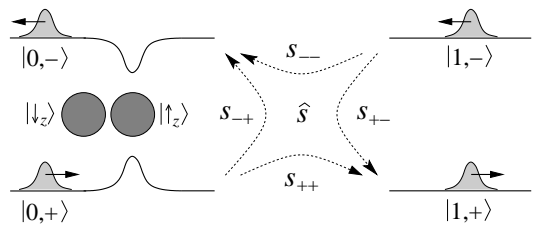}
\caption{An electron in an incoming superposition state $a|0,+\rangle +
b|1,-\rangle$ undergoes scattering in a quantum point contact (a QPC, here
implemented in a Quantum Hall bar device) described by the scattering matrix
$\hat{s}$ with symmetric scattering coefficients $s_{ij}$, $i,j = \pm$. The
states $|0,\pm\rangle$ interact with a double dot residing in one of the
operational states $|\!\!\uparrow\rangle$ or $|\!\!\downarrow \rangle$ which
are superpositions of the physical states $|\!\!\uparrow_z\rangle$ and
$|\!\!\downarrow_z \rangle$. The interaction between the double-dot and the
scattering states $|0,\pm\rangle$ generates the conditional (on
$|\!\!\Uparrow\rangle$) $\pi$-phase shift $\hat{u}$.  Combining this operation
with the scattering of the electron at the QPC and a $\pi/2$-rotation of the
double-dot, see text, swaps the electron and double-dot states.}
\label{fig:dd}
\end{center}
\end{figure}

Consider an electron wave packet (our system) which propagates through the
edge states of an Integer Quantum Hall bar device and which scatters at a
quantum point contact (QPC), see Fig.\ \ref{fig:dd}.  The electron arrives at
the QPC through the incoming edge channels $|0,+\rangle$ and $|1,-\rangle$ and
scatters into the outgoing leads $|0,-\rangle$ and $|1,+\rangle$. It is
convenient to use the pseudo-spin notation $|0,+\rangle =
|\!\!\Uparrow\rangle$ and $|1,-\rangle = |\!\!\Downarrow\rangle$ for the
incoming states and the same for the outgoing ones, $|0,-\rangle =
|\!\!\Uparrow\rangle$ and $|1,+\rangle = |\!\!\Downarrow\rangle$, while keeping in
mind to switch meaning when going from incoming to outgoing leads.  The
elastic scattering process is described by a symmetric scattering matrix
$\hat{s}(\theta, \eta)$, see Eq.\ (\ref{eq:scmat}). The micro-environment
interacting with the scattering electron is given by a double-dot, replacing
the spin-environment in the previous section.  We use the spin notation
$|\!\!\uparrow_z\rangle$ ($|\!\!\downarrow_z\rangle$) to describe the
double-dot's semi-classical state with a localized charge in the right (left)
dot of the qubit.

We place the charge qubit to the left of the QPC such that it interacts
capacitively and symmetrically with the electron propagating in either of the
two left leads $|\!\!\Uparrow\rangle$ (i.e., both $|0,\pm\rangle$), while
leaving the right leads $|\!\!\Downarrow\rangle$ invariant.  Our task is to
find an interaction and an operational basis $|\!\!\uparrow\rangle$ and
$|\!\!\downarrow\rangle$ that satisfy the conditions \eqref{eq:untr1} and
\eqref{eq:untr2}. The latter tell, that $\hat{u}_3$ shall add a phase
difference of $\pi$ to the operational states and $\hat{u}_1$ shall exchange
them.  The electron--qubit interaction acts on the physical qubit states
$|\!\!\uparrow_z\rangle$ and $|\!\!\downarrow_z\rangle$, provided the electron
passes the double-dot in a $|\!\!\Uparrow\rangle$ state. We then can use the
interaction to generate the relative phase shift $\pi$, e.g., by placing the
two dots in a symmetric manner between the left incoming and outgoing wires
and choosing a geometry that couples one of the dots more strongly to the
leads, see Fig.\ \ref{fig:dd}.  Hence, we demand that the electron--qubit
interaction shall generate the unitary transformation (a rotation around the
$z$-axis)
\begin{align} \label{eq:unitr3}
   \hat{U}|\!\!\Uparrow \rangle \!\otimes |\!\!\uparrow_z\rangle
    &= \phantom{+} e^{i\phi}|\!\!\Uparrow\rangle \!\otimes
   |\!\!\uparrow_z\rangle, \\ \nonumber
   \hat{U}|\!\!\Uparrow\rangle \!\otimes |\!\!\downarrow_z\rangle
   &= -e^{i\phi}|\!\!\Uparrow\rangle \!\otimes |\!\! \downarrow_z\rangle,
   \\ \nonumber
   \hat{U}|\!\!\Downarrow \rangle \!\otimes |\!\!\uparrow_z\rangle
    &= |\!\!\Downarrow\rangle \!\otimes |\!\!\uparrow_z\rangle, \\
    \nonumber
   \hat{U}|\!\!\Downarrow\rangle \!\otimes |\!\!\downarrow_z\rangle
    &= |\!\!\Downarrow\rangle \!\otimes |\!\! \downarrow_z\rangle.
\end{align}
We have used a capital letter $\hat{U}$ to denote the operator acting on the
grand system. The action of $\hat{U}$ on the joint system then corresponds to
a controlled (relative) phase shift.

However, according to \eqref{eq:untr2} it is the operational states
$|\!\!\uparrow\rangle$ and $|\!\!\downarrow\rangle$ that should pick up this
relative phase; furthermore, the latter should be flipped by the operation
$\hat{u}_1$. Hence, we should introduce a $\pi/2$ rotation that transforms
between planar operational states $|\!\!\uparrow\rangle$ and
$|\!\!\downarrow\rangle$ and axial physical states $|\!\!\uparrow_z\rangle$
and $|\!\!\downarrow_z\rangle$.

In order to rotate the qubit state, we assume a tunable double-dot described
by the free Hamiltonian $\hat{H}_0 = \epsilon_0 \, [\,{|\!\!\uparrow_z
\rangle} {\langle \uparrow_z \!\!|} + {|\!\!\downarrow_z\rangle} {\langle
\downarrow_z\!\!|}\,] - [\,\Delta(t)\,{|\!\!  \uparrow_z\rangle} {\langle
\downarrow_z\!\!|} + \mathrm{h.c.}\,]$, i.e., the upper and lower dots of the
qubit have equal energies $\epsilon_0$ (we assume $\epsilon_0 = 0$ from now
on) and the tunneling amplitude $\Delta(t) \equiv |\Delta(t)|e^{i\varphi}$
between the dots can be dynamically changed. Assuming a finite but constant
$\Delta(t)$, the evolution of the qubit is described by the unitary operator
$\hat{u}(\tau) = \exp(-i\hat{H}_0\tau/\hbar)$. In particular, we consider the
specific unitary transformation $\hat{u}_{\scriptscriptstyle 1/4} \equiv
\hat{u}(\tau_{\scriptscriptstyle 1/4})$ for a fixed time interval
$\tau_{\scriptscriptstyle 1/4}=\hbar\pi/4|\Delta|$,
\begin{equation}\label{eq:u14}
      \hat{u}_{\scriptscriptstyle 1/4} = \frac1{\sqrt{2}}\left(
      \begin{array}{cc}
      1&ie^{i\varphi}\\ ie^{-i\varphi}&1
      \end{array}\right),
\end{equation}
that rotates the physical states $|\!\!\uparrow_z\rangle$ and
$|\!\!\downarrow_z\rangle$ into the $xy$-plane. Defining the operational
states 
\begin{align}\label{eq:op-states}
   |\!\!\uparrow\rangle &= \hat{u}_{\scriptscriptstyle 1/4}|\!\!\uparrow_z\rangle,\\
   \nonumber
  |\!\!\downarrow\rangle &= \hat{u}_{\scriptscriptstyle 1/4} |\!\!\downarrow_z\rangle,
\end{align}
the electron--qubit interaction in Eq.\ (\ref{eq:unitr3}) acts on the demon
states with the following unitary
\begin{eqnarray}\label{eq:u1}
   &&\hat{u}\, |\!\!\uparrow\rangle = - i e^{i(\phi - \varphi)} 
   |\!\!\downarrow\rangle, \\ \nonumber
   &&\hat{u}\, |\!\!\downarrow\rangle = \phantom{+}ie^{i(\phi+\varphi)}
   |\!\!\uparrow\rangle,
\end{eqnarray}
which is equivalent to the action of the $\hat{u}_1$ operator, see Eq.\
(\ref{eq:untr1}) with $\alpha = \phi - \varphi - \pi/2$ and $\beta =
\phi + \varphi + \pi/2$, i.e., $\hat{u} = \hat{u}_1 (\alpha,\beta)$. On the
other hand, we can implement the simple phase shift operation of $\hat{u}_3$,
see Eq.\ (\ref{eq:untr2}), by rotating the operational state back to the
physical state and let the interaction act with the transformation $\hat{u}$
once more.  The transformation between the operational $|\!\!\uparrow\rangle,
|\!\!\downarrow\rangle$ and the physical $|\!\!\uparrow_z\rangle,
|\!\!\downarrow_z\rangle$ basis states is given by
$\hat{u}_{\scriptscriptstyle 1/4}$, the unitary gate in Eq.\ \eqref{eq:u14}
which is written in the physical $|\!\!\uparrow_z\rangle,
|\!\!\downarrow_z\rangle$ basis. Expressing the operational basis through the
physical one, $|\!\!\uparrow\rangle = \bigl[ {|\!\!\uparrow_z\rangle} + i
e^{-i\varphi} {|\!\!\downarrow_z\rangle} \bigr]/\sqrt{2}$ and
$|\!\!\downarrow\rangle = \bigl[i e^{i\varphi} |\!\!\uparrow_z\rangle +
|\!\!\downarrow_z\rangle \bigr]/\sqrt{2}$, one finds that
$\hat{u}_{\scriptscriptstyle 1/4}|\!\!\uparrow\rangle =
ie^{-i\varphi}|\!\!\downarrow_z\rangle$ and $\hat{u}_{\scriptscriptstyle
1/4}|\!\!\downarrow\rangle = ie^{i\varphi}|\!\!\uparrow_z\rangle$, and hence
\begin{eqnarray}\label{eq:unitr4}
   &&\hat{u}_{\scriptscriptstyle 1/4}^\dagger \hat{u} \,
   \hat{u}_{\scriptscriptstyle 1/4} |\!\!\uparrow\rangle  
   = - e^{i\phi} |\!\!\uparrow\rangle, \\ \nonumber
   &&\hat{u}_{\scriptscriptstyle 1/4}^\dagger \hat{u} \,
   \hat{u}_{\scriptscriptstyle 1/4} |\!\!\downarrow\rangle  
   = \phantom{+} e^{i\phi} |\!\!\downarrow\rangle,
\end{eqnarray}
i.e., $\hat{u}_{\scriptscriptstyle 1/4}^\dagger \hat{u}\,
\hat{u}_{\scriptscriptstyle 1/4} = \hat{u}_3$. We conclude that in the present
setup, the two non-commuting unitary operations $\hat{u}_1$ and $\hat{u}_3$
can be realized by letting the electron and the double-dot interact twice in
the same manner, once in the incoming and a second time in the outgoing lead,
provided that the qubit is rotated by $\hat{u}_{\scriptscriptstyle 1/4}$
between the two interaction events (note that the last single-qubit rotation
$\hat{u}_{\scriptscriptstyle 1/4}^\dagger$ can be dropped).  For an electron
wavepacket with a finite width, the interaction with the double-dot has to be
separated in time from the subsequent scattering at the QPC. The evolution of
the overall system then can be separated into four steps, i) preparation:
starting with a physical (localized) state $|\!\!\uparrow_z\rangle$ or
$|\!\!\downarrow_z\rangle$, the tunneling $\Delta$ is switched on during the
time $\tau_{\scriptscriptstyle 1/4}$ in order to generate the operational
states $|\!\!\uparrow\rangle$ and $|\!\!\downarrow\rangle$. ii) first
interaction: the electron interacts with the double-dot in the incoming lead,
thereby generating the transformation $\hat{u} = \hat{u}_1$. iii) single-qubit
rotations: while the electron undergoes scattering at the QPC, the double-dot
is rotated by $\hat{u}_{\scriptscriptstyle 1/4}$ (by again switching on the
tunneling $\Delta$ during the time $\tau_{\scriptscriptstyle 1/4}$) as part of
the $\hat{u}_3$ operation. iv) second interaction: the electron interacts with
the double-dot in the outgoing lead as part of the $\hat{u}_3$ transformation.
As we can drop the last qubit rotation $\hat{u}_{\scriptscriptstyle
1/4}^\dagger$ this completes the sequence. The above evolution transforms any
initial electron (or flying qubit) state $|\phi_0\rangle =
a\,|\!\!\Uparrow\rangle + b\,|\!\!\Downarrow\rangle$ into the pure planar
states $|\!\!\Uparrow_{xy} \rangle$ or $|\!\!\Downarrow_{xy} \rangle$ as given
in Eq.\ \eqref{eq:Upxy}.  Keeping $\Delta = 0$ during the process (except for
the time intervals during the demon rotations) the above evolution occurs
without energy exchange between the electron and the double dot. Thus, the
described protocol produces the same energy-conserving non-unital quantum
channel as described in the previous section but requires an external
manipulation of the demon qubit between the two interaction events.

\begin{figure}[h]
\begin{center}
\includegraphics[width=7truecm]{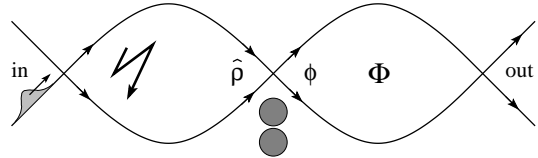}
\caption{Schematic version of a double Mach-Zehnder Interferometer with
quantum demon. The incoming wave-packet is decohered (arrow) in the first loop
and restored to a pure state $|\phi\rangle$ by the action of the demon
(double-dot). The second loop tests the coherence of the wave function by
shifting the relative phase in the two arms with the help of a tunable flux
$\Phi$. The appearance of Aharonov-Bohm oscillations in the outgoing arms
testifies for the successful action of the demon.}
\label{fig:ahar}
\end{center}
\end{figure}

The functionality of the demon's purification action can be experimentally
demonstrated in a mesoscopic quantum interference setting as sketched in Fig.\
\ref{fig:ahar}.  To this end, we consider two consecutive Mach-Zehnder
interferometers where the electron first looses its coherence within the first
loop through a dephasing process, e.g., via entanglement with another qubit.
The electron then arrives at the intermediate QPC in a mixed state (mani\-fest
after tracing out the entangled qubit). The demon-enhanced QPC operating with
a pure demon qubit then restores the purity of the electron state, what can be
observed in the resurrection of Aharonov-Bohm oscillations in the outgoing
leads of the second interferometer as the flux $\Phi$ penetrating the
second loop is varied.  Furthermore, replacing the demon's qubits by mixed
states, these Aharonov-Bohm oscillations can be tuned with a visibility
changing between one and zero. 

The Mach-Zehnder setup motivates a yet other formulation of the demon's
functionality. Injecting an electron (1) into the first loop, entangling it
with a dephasing qubit (2), and performing the demon's action involving the
demon qubit (3) swaps the entanglement between (1) and (2) to the entanglement
between (2) and (3), letting (1) further evolve in a pure state. Note that
while a direct measurement of entropy reduction is difficult, the observation
of increased visibility in a Mach-Zehnder device due to improved coherence
offers an attractive substitute.

Also, it is interesting to compare the different functionalities of classical
and quantum Maxwell demons.  In the above setup, the action of the quantum
demon generates a prescribed electronic quantum state in the second
interferometer.  A corresponding classical demon\cite{szillard:29} achieving
this task would first measure the electron's state in the first interferometer
and, depending on the measurement outcome (upper or lower lead), adjust the
scattering matrix $\hat{s}$ in such a way as to produce the desired quantum
state in the second interferometer through the scattering process.  The notion
`classical' then refers to the demon's property to acquire classical
information on the system.  The functionalities of the two schemes are very
different: the classical Maxwell demon interrupts the quantum evolution of the
system by measuring its quantum state. The obtained classical information is
subsequently used in a conditional manipulation of the system.  The action of
the quantum Maxwell demon is conceptually different: the system's quantum
evolution is never interrupted and the evolution of the joint system--demon
setup is autonomous.  Furthermore, although the quantum demon acquires the
initially unknown system state, see Eq.\ (\ref{eq:swap1}), no use is made of
this information in the process; in fact, this information is discarded after
the pSWAP operation. Instead, it is the known quantum state stored in the
demon and later deposited into the system that is relevant for the
functionality of the process. Hence, while the classical Maxwell demon {\it
acquires} some classical information about the system's state, our quantum
Maxwell demon {\it places} some known quantum state into the system.

In this context, we also refer to the analysis in Ref.\
[\onlinecite{horowitz:14}] where the action of a classical Maxwell demon was
included in the entropy balance of the system by introducing the {\it mutual
information} $I_{{\cal S};{\cal D}} \geq 0$ quantifying the information stored
in the demon about the system. The Second Law as applied to the system alone
then has to be extended \cite{horowitz:14} to include this mutual information,
$\Delta S_{\cal S} + I_{{\cal S};{\cal D}} \geq 0$.  The inequality tells that
the information on the system learnt by the demon can be used to reduce the
system's entropy by the maximal amount $\Delta S_{\cal S} = -I_{{\cal S};{\cal
D}}$. It is important to distinguish this classical type of entropy decrease
from our quantum process. The literal application of the analysis in Ref.\
[\onlinecite{horowitz:14}] to our quantum setup cannot explain its entropy
decrease in the scattering process: since the flying and demon qubits start
out and end up in a product state before and after the scattering, we have
$I_{{\cal S};{\cal D}} = 0$ and hence the classical expectation is that
$\Delta S_{\cal S} \geq 0$, in obvious conflict with our entropy decrease due
to the pSWAP.

\section{Quantum circuit description}\label{sec:qcd}

\begin{figure}[htbp]
\begin{center}
\includegraphics[width=8truecm]{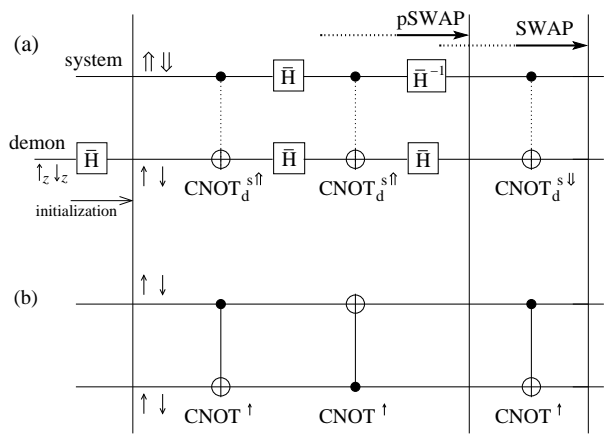}
\caption{(a) Quantum circuit describing the demon action as a partial SWAP or
pSWAP and its extension to a full SWAP operation. The dotted lines on the CNOT
operations indicate that no energy is exchanged between the system and demon
qubits. (b) Minimal pSWAP and SWAP operations using two and three CNOT
operations with exchanged controller\cite{schuch:03}.}
\label{fig:pSWAP}
\end{center}
\end{figure}

The demon's action can be conveniently formalized in the quantum circuit
language. Indeed, each stage of the system--demon evolution can be described
by a corresponding quantum gate. We choose the product basis
$\{{|\!\!\Uparrow\rangle}, {|\!\!\Downarrow\rangle}\} \otimes
\{{|\!\!\uparrow\rangle}, {|\!\!\downarrow\rangle}\}$ as our re\-presentation of
the initial joint system state and the product basis $\{{|\!\!\Uparrow\rangle},
{|\!\!\Downarrow\rangle}\} \otimes \{{|\!\!\uparrow\rangle},
{|\!\!\downarrow\rangle}\}$ as our representation of the final joint system
state and use lower case (capital) letters to denote operators acting on the
individual (joint) system.  Below we consider a specific situation with the
interaction phase $\phi = 0$ and the tunneling phase $\varphi = -\pi/2$,
implying $\alpha = 0$ and $\beta = 0$.  Then the unitary rotation $\hat{u}$
induced by the electron--qubit interaction as described in Eq.\ (\ref{eq:u1})
corresponds to a CNOT quantum gate when extended to the joint electron--demon
system, $\hat{U} = \mathrm{CNOT}^\mathrm{s \Uparrow}_\mathrm{d}$, where the
electron or system state serves as the control, $\hat{U}(|\!\!\Uparrow \rangle
\otimes |\!\!\uparrow\rangle) = |\!\!\Uparrow \rangle \otimes
|\!\!\downarrow\rangle$ and $\hat{U}(|\!\!\Uparrow \rangle \otimes
|\!\!\downarrow\rangle) = |\!\!\Uparrow \rangle \otimes |\!\!\uparrow\rangle$,
while $\hat{U}(|\!\!\Downarrow\rangle \otimes |\!\!\uparrow\rangle) =
|\!\!\Downarrow\rangle \otimes |\!\!\uparrow\rangle$ and
$\hat{U}(|\!\!\Downarrow\rangle \otimes |\!\!\downarrow\rangle) =
|\!\!\Downarrow\rangle \otimes |\!\!\downarrow\rangle$.  In between the
interaction events the evolution is described by two single-qubit rotations,
one on the electron and another on the qubit. In the demon's operating basis,
the qubit rotation $\hat{u}_{\scriptscriptstyle 1/4}$ corresponds to a
Hadamard-type gate, $\hat{u}_{\scriptscriptstyle 1/4} = \sigma_z \mathrm{H}
\equiv \bar{\mathrm{H}}$ with $\mathrm{H}$ the standard Hadamard operation,
$\hat{u}_{\scriptscriptstyle 1/4}|\!\!\uparrow\rangle =
\bigl(|\!\!\uparrow\rangle - |\!\!\downarrow\rangle \bigr) /\sqrt{2}$ and
$\hat{u}_{\scriptscriptstyle 1/4}|\!\!\downarrow\rangle = \bigl(
{|\!\!\uparrow\rangle} + |\!\!\downarrow\rangle \bigr) / \sqrt{2}$.  The
electron state is transformed according to the scattering matrix
$\hat{s}(\theta,\eta)$ of the QPC; for the specific choice $\theta = 0$ and
$\eta = \pi$ it can be represented by our Hadamard-type gate as well,
$\hat{s}(0,\pi) = \bar{\mathrm{H}}$. Hence, the overall unitary transformation
of the demon's purification protocol (expressed in the basis
$\{|\!\!\Uparrow\rangle \!\otimes |\!\!\uparrow\rangle, |\!\!\Uparrow\rangle
\!\otimes {|\!\!\downarrow\rangle}, {|\!\!\Downarrow\rangle} \!\otimes
|\!\!\uparrow\rangle, |\!\!\Downarrow\rangle \!\otimes |\!\!\downarrow\rangle
\}$) is
\begin{align} \label{eq:U_D}
  \hat{U}_\mathcal{D}
   &= \mathrm{CNOT}^\mathrm{s\Uparrow}_\mathrm{d}\cdot
   \bigl[\bar{\mathrm{H}}_\mathrm{s} \otimes \bar{\mathrm{H}}_\mathrm{d}\bigr]
    \cdot \mathrm{CNOT}^\mathrm{s\Uparrow}_\mathrm{d}  \\ \nonumber
   &\qquad= \frac{1}{2}
   \left( \begin{array}{rrrr}
   1 & \!\!-1 & \!\!-1 & 1 \\ \!\!1 & 1 & \!\!1 & 1 \\
   -1 & -1 & 1 & 1 \\ \!\!-1 & \!\!1 & -1 & 1
   \end{array} \right).
\end{align}
The simplest explicit form of the $\hat{U}_\mathcal{D}$ matrix is obtained
when performing two additional Hadamard-type gates at the output of the
circuit, $\hat{V}_\mathcal{D} = \bigl[ \bar{\mathrm{H}}_\mathrm{s}^{-1}
\otimes \bar{\mathrm{H}}_\mathrm{d}\bigr] \cdot \hat{U}_\mathcal{D}$ (see
Fig.\ \ref{fig:pSWAP} for the corresponding quantum circuit diagram; note that
the last rotation $\bar{H}_\mathrm{d}$ is not the rotation
$\hat{u}^\dagger_{\scriptscriptstyle 1/4}$ completing the operation
$\hat{u}_3$ in (\ref{eq:unitr4})),
\begin{align} \nonumber
  \hat{V}_\mathcal{D}
   &= \bigl[\bar{\mathrm{H}}^{-1}_\mathrm{s}\otimes 
   \bar{\mathrm{H}}_\mathrm{d}\bigr]
   \cdot \mathrm{CNOT}^\mathrm{s\Uparrow}_\mathrm{d}\cdot
   \bigl[\bar{\mathrm{H}}_\mathrm{s} \otimes 
   \bar{\mathrm{H}}_\mathrm{d}\bigr]
    \cdot \mathrm{CNOT}^\mathrm{s\Uparrow}_\mathrm{d} \\
   &\qquad\qquad \qquad
   = \left( \begin{array}{cccc} 1&0&0&0\\0&0&1&0\\0&0&0&1\\0&1&0&0
  \end{array} \right).\label{eq:V_D}
\end{align}
The operations $\hat{U}_\mathcal{D}$ and $\hat{V}_\mathcal{D}$ both describe
the pSWAP operation of the demon and differ only by a final change of basis.

The operation $\hat{V}_\mathcal{D}$ acts as a SWAP only on one pair of basis
states,
\begin{align}\label{eq:ac-VD1}
      \hat{V}_\mathcal{D}\> |\!\!\Uparrow\rangle \otimes |\!\!\uparrow\rangle
      &= |\!\!\Uparrow\rangle \otimes |\!\!\uparrow\rangle, \\ \nonumber
      \hat{V}_\mathcal{D}\> |\!\!\Downarrow\rangle \otimes |\!\!\uparrow\rangle
      &= |\!\!\Uparrow\rangle \otimes |\!\!\downarrow\rangle,
\end{align}
while the second pair is swapped up to a NOT operation on the demon qubit,
\begin{align}\label{eq:ac-VD2}
   \hat{V}_\mathcal{D}\> |\!\!\Uparrow\rangle \otimes |\!\!\downarrow\rangle
   &= |\!\!\Downarrow\rangle \otimes |\!\!\downarrow\rangle, \\ \nonumber
   \hat{V}_\mathcal{D}\> |\!\!\Downarrow\rangle \otimes |\!\!\downarrow\rangle
   &= |\!\!\Downarrow\rangle \otimes |\!\!\uparrow\rangle,
\end{align}
hence the name partial SWAP or pSWAP. Summarizing, the demon's operation
$\hat{V}_\mathcal{D}$ swaps the system- and demon qubit states (up to an
independent unitary rotation of each subsystem) if the demon qubit was
initially prepared in a pure operational state $|\!\!\uparrow\rangle = (1,0)$
or $|\!\!\downarrow\rangle = (0,1)$, $\hat{V}_\mathcal{D} [(a,b)\otimes (1,0)]
= (1,0) \otimes (a,b)$ and $\hat{V}_\mathcal{D} [(a,b)\otimes (0,1)] = (0,1)
\otimes (b,a)$.  However, for an arbitrary initial state of the
qubit, the circuit does not operate as a SWAP gate, $\hat{V}_\mathcal{D}
[(a,b)\otimes (\alpha,\beta)] \neq (\alpha,\beta) \otimes (a,b)$.  

In order to arrive at a full SWAP operation, we have to add another CNOT
operation which acts on the demon when the controlling system is in the
$|\!\Downarrow\rangle$ state,
\begin{align}\label{eq:SWAP}
   \mathrm{SWAP} = \mathrm{CNOT}_\mathrm{d}^{\,\mathrm{s}\scriptscriptstyle
   \Downarrow} \cdot \hat{V}_\mathcal{D}
   = \left( \begin{array}{cccc}
   1 & 0 & 0 & 0 \\ 0 & 0 & 1 & 0 \\ 0 & 1 & 0 & 0 \\ 0 & 0 & 0 & 1
   \end{array} \right).
\end{align}
However, such a gate would require a more sophisticated interaction that is
not easily realized within our physical setting, while an implementation using
spins and proper NMR sequences seems more promising \cite{lloyd:97}.  The
circuit dia\-gram for the pSWAP and its extension to the SWAP operation are
shown in Fig.\ \ref{fig:pSWAP}.

\section{Quantum thermodynamic engine} \label{sec:qte}

The energy-conserving pSWAP operation invites the design of a
quantum-thermodynamic engine \cite{lloyd:97,quan:06,delacruz:14} which
transforms heat from a single reservoir into work (or ordered energy) by
making use of quantum purity provided by a demon with energy-degenerate
quantum states.  The specific design of our machine naturally separates its
operation into two cycles with no energy transfer across, one cycle providing
energy or work from a single thermal reservoir, while the purification of the
demon's qubits can be handled in a separate `entropy cycle' away from the
engine.  This contrasts with the operation of a classical machine where the
maximal extractable work is given by the Helmholtz free energy $\delta F =
\delta U -T\delta S$ of the thermodynamic process\cite{procaccia:76} (the same
holds for a class of quantum engines\cite{skrzypczyk:14}), binding the energy
and entropy flows in the process.

\begin{figure}[htbp]
\begin{center}
\includegraphics[width=7truecm]{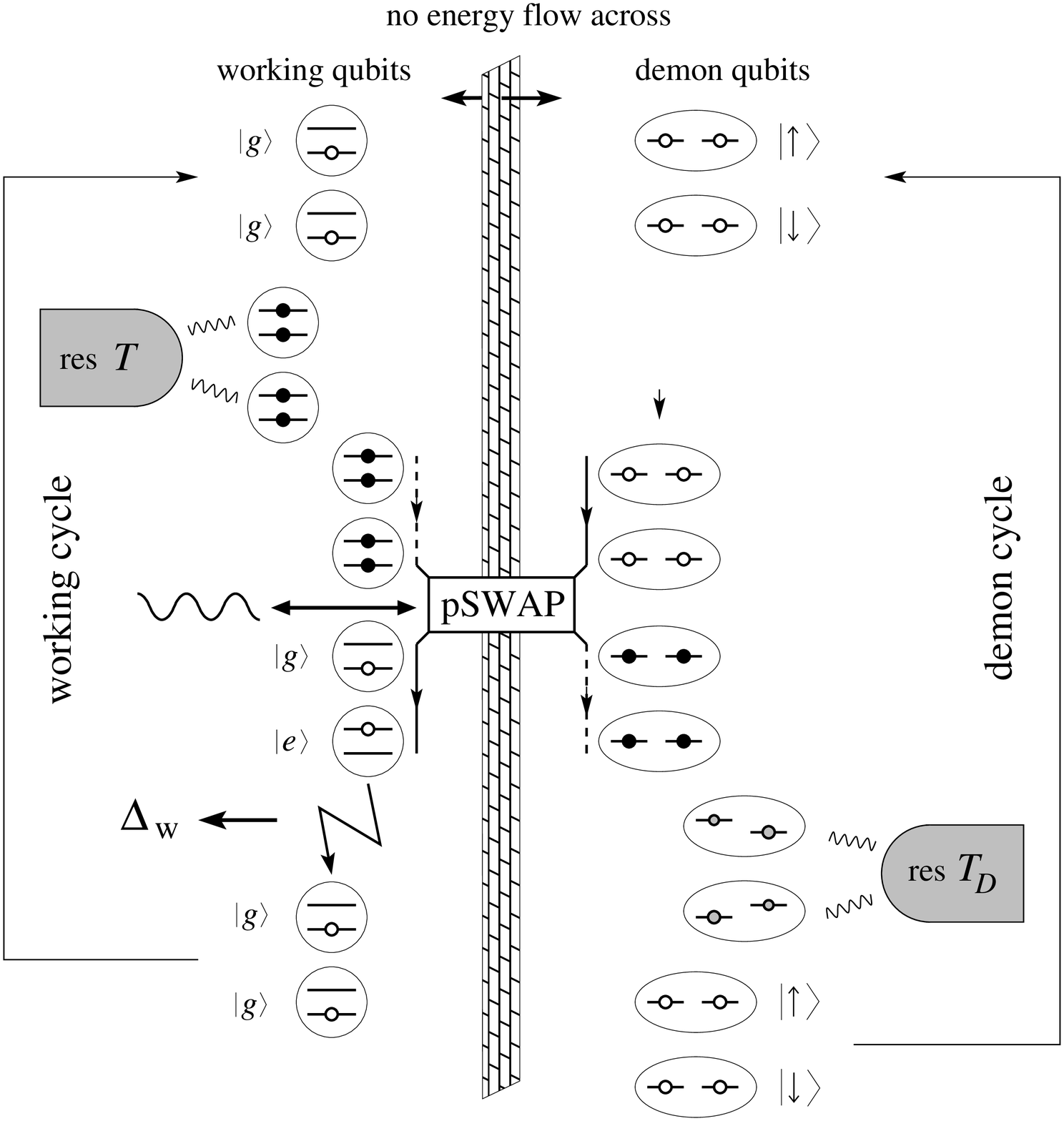}
\caption{Quantum thermodynamic engine with two energeti\-cally separated
cycles. The working cycle (left) starts with two working qubits (wits) in the
pure (empty dots) ground state $|g\rangle$. The contact with the thermal
reservoir excites them into a mixed (solid dots) state. After the demon's
swap, the wits have taken over the pure states from the demon qubits (dits),
$|\!\!\uparrow\rangle \to |e\rangle$ and $|\!\!\downarrow\rangle \to
|g\rangle$. The energy $\Delta_\mathrm{w}$ of the excited wit can be used for
work extraction. Different from the pSWAP in Fig.\ \ref{fig:setup}, the
present pSWAP has an energy context involving a classical field (wavy line).
The entropy cycle (right) starts with two pure dits in states
$|\!\!\uparrow\rangle$ and $|\!\!\downarrow\rangle$ which are `consumed' by
the wits and then reprepared for the next cycle.}
\label{fig:machine}
\end{center}
\end{figure}

The engine consists of two (identical) working qubits (two two-level systems)
with fixed energy-level spacing $\Delta_\mathrm{w}$ and a thermal reservoir at
the temperature $T$ enhanced with a quantum demon providing pure qubits, see
Fig.\ \ref{fig:machine}. We first analyse an idealized working cycle and
discuss issues related with practical implementations later on.

The two working qubits (wits), assuming the role of the system qubits, start
out in their ground states $|g\rangle \leftrightarrow |\!\!\Downarrow\rangle$
and then are placed into thermal contact with a macroscopic heat reservoir at
a temperature $T$.  After the isochoric thermalisation (i.e., the wit's
spectra remain unchanged), both wits are detached from the heat reservoir;
their states are described by the density matrices $\hat \rho_\mathrm{w} =
Z^{-1} \exp(-\beta \hat{H}_\mathrm{w}) \equiv p_g|g\rangle \langle g| + p_e
|e\rangle \langle e|$, where $\hat{H}_\mathrm{w} = \Delta_\mathrm{w} |e\rangle
\langle e|$ is the free Hamiltonian of the wits (with $|e\rangle
\leftrightarrow |\!\!\Uparrow\rangle$), $\beta = 1/T$ is inverse temperature
with the Boltzmann constant set to unity, $p_g = 1/Z$, $p_e =
e^{-\beta\Delta_\mathrm{w}}/Z$, and $Z = 1 + e^{-\beta\Delta_\mathrm{w}}$.

Next, a pSWAP operation with a pair of demon qubits (dits) prepared in pure
operational energy-degenerate states $\hat{r}_\uparrow = |\!\!\uparrow\rangle
\langle \uparrow\!\!|$  and $\hat{r}_\downarrow = |\!\!\downarrow\rangle
\langle \downarrow\!\!|$ exchanges wit and dit states,
\begin{align}\label{eq:wswap}
   \mathrm{pSWAP} &= \bigl[ \hat{u}_{\scriptscriptstyle 1/4} \otimes 
   \mathds{1}\bigr]\cdot \mathrm{CNOT}_\mathrm{d}^\mathrm{w\Uparrow} \\
   \nonumber
   &\qquad \qquad \cdot \bigl[ \hat{u}_{\scriptscriptstyle 1/4} \otimes 
   \bar{\mathrm{H}}_\mathrm{d}\bigr] \cdot 
   \mathrm{CNOT}_\mathrm{d}^\mathrm{w\Uparrow},
\end{align}
where $\hat{u}_{\scriptscriptstyle 1/4}$ is a Hadamard-type transformation of
the working qubit, see Eq.\ (\ref{eq:u14}).  E.g., assuming an interaction
between the wits and dits that generates a conditional $\pi$-shift as
described in Eq.\ \eqref{eq:unitr3} with $\phi = 0$, we choose appropriate
operational dit states $|\!\!\uparrow\rangle$ and $|\!\!\downarrow\rangle$
according to Eq.\ (\ref{eq:op-states}); this generates a
CNOT$_\mathrm{d}^\mathrm{w\Uparrow}$ operation that flips the dit when the wit
is in the excited state.  After the pSWAP operation, the two wit--dit systems
reside in the states (we assume phases $\theta = \phi = 0$ and $\eta = \pi/2 -
\varphi$)
\begin{eqnarray}
   &&\mathrm{pSWAP} \bigl[ \hat\rho_\mathrm{w} \!\otimes \hat{r}_\uparrow \bigr]
   = \hat{\rho}_g \!\otimes \hat{r}_+,\>\>
   \\
   &&\mathrm{pSWAP} \bigl[ \hat\rho_\mathrm{w} \!\otimes \hat{r}_\downarrow \bigr]
   = \hat{\rho}_e \!\otimes \hat{r}_-,\>\>
\end{eqnarray}
where $\hat{\rho}_g = |g\rangle \langle g|$, $\hat{\rho}_e = |e\rangle \langle
e|$ and $\hat{r}_+ = p_g |\!\!\uparrow\rangle \langle \uparrow\!\!| + p_e
{|\!\!\downarrow\rangle}{\langle\downarrow\!\!|}$, $\hat{r}_- = p_g
|\!\!\downarrow\rangle \langle\downarrow\!\!| + p_e |\!\!\uparrow\rangle
\langle \uparrow\!\!|$ are final mixed states of the demon qubits, see Eq.\
(\ref{eq:density1}). The net average energy cost of this pair of
pSWAP operations is given by the difference of the final and initial average
energy of the working qubits,
\begin{equation}\label{eq:W-}
   W_- =\Delta_\mathrm{w} - 2p_e \Delta_\mathrm{w} = \Delta_\mathrm{w}\,
   \frac{1 - e^{-\beta\Delta_\mathrm{w}}}{1 + e^{-\beta\Delta_\mathrm{w}}}.
\end{equation}
The pSWAP operation deterministically pushes the first wit into the ground
state and the second wit into its excited state. The latter now stores an
ordered energy (or work) $W_+ = \Delta_\mathrm{w}$ which can be extracted,
e.g., with the help of a half-Rabi pulse $\hat{u}_{\scriptscriptstyle 1/2}
|e\rangle = \hat{u}_{\scriptscriptstyle 1/4} \hat{u}_{\scriptscriptstyle 1/4}
|e\rangle = i e^{-i\varphi} |g\rangle$ of coherent radiation.  Calculating the
energy balance of the working cycle, one finds that this cycle operates with
unit efficiency $\eta  = 1$, as all of the initial heat $Q = 2p_e
\Delta_\mathrm{w}$ stored in the two wits is transformed into net useful work
$W_\mathrm{out} = W_+ - W_- = Q$ in a deterministic fashion.  This working
cycle then consumes the purity provided by the two energy-degenerate demon
qubits and we will discuss the cost of this consumption below.

In the sequence Eq.~(\ref{eq:wswap}), the two transformations
$\hat{u}_{\scriptscriptstyle 1/4}$  on the wits require an external supply of
work \cite{aberg:2014} since they transform between the pure-energy wit
mixtures of $|g\rangle$ and $|e\rangle$ and the balanced superposition states
$|s_1\rangle = [ie^{i\varphi}|e\rangle +|g\rangle]/\sqrt{2}$ and $|s_2\rangle
= [|e\rangle + ie^{-i\varphi}|g\rangle]/\sqrt{2}$. This process requires an
external work $\Delta_\mathrm{w} /2$ or $-\Delta_\mathrm{w}/2$ to be provided
to or extracted from the working qubits (the final
$\hat{u}_{\scriptscriptstyle 1/4}$ operation transforms the superpositions
$|s_1 \rangle$ and $|s_2\rangle$ into the pure-energy states $|g\rangle$ and
$|e\rangle$).  For a pair of wits in fully chaotic states, the combination of
these operations does not involve a net {\it average} energy transfer between
the field and the qubits and hence does not require an external flow of
energy.  This (at least on average) energy conservation is due to our use of
two demon qubits residing in opposite states $|\!\!\uparrow\rangle$ and
$|\!\!\downarrow\rangle$: while the demon qubit in state
$|\!\!\downarrow\rangle$ always generates a wit in the excited state
$|e\rangle$, the dit in state $|\!\!\uparrow\rangle$ pushes the wit to the
ground state, hence, on average an equal amount of energy is extracted from or
deposited into the classical field. In the case where we were to use only one
wit--dit pair with the dit in state $|\!\!\downarrow\rangle$, only half of the
energy gained in the working cycle would originate from the thermal bath,
while the other half would have to be provided by the classical field.  As the
thermally excited wits on average contain less than $\Delta_\mathrm{w}$ on
energy, the difference $W_-$ to the final energy $\Delta_\mathrm{w}$ after the
pSWAP operation has to be provided by the classical field, see Eq.\
\eqref{eq:W-}.

The $\hat{u}_{\scriptscriptstyle 1/4}$ operations changing the wit's energies
in the pSWAP can be realized by letting the wits coherently exchange
energy with a classical field in resonance with their energy spacing
$\Delta_\mathrm{w}$.  Assuming a dipolar interaction of the wits (featuring a
dipole matrix element $d$) with a coherent field of strength $E_0$, see Ref.\
[\onlinecite{mandel-wolf:95}], this operation can be described by the
Hamiltonian $\hat{H}_\mathrm{dip} = -E_0\, \bigl(d\, |e\rangle \langle g| +
d^*\,|g\rangle \langle e|\, \bigr) \cos(\Delta_\mathrm{w} t)$. Within a
rotating-wave approximation, the field-driven evolution of the working qubits
involves Rabi-oscillations with a frequency $\Omega = E_0|d|/2$. Keeping the
wit during the time $\tau_{\scriptscriptstyle 1/4} = \pi/4\Omega$ in resonance
with the field, the corresponding unitary transformation of the basis states
$\{ |e\rangle, |g\rangle\}$ is given by Eq.\ (\ref{eq:u14}) with the phase
$\varphi = \arg(d)$.

The above feature introduces a distinct difference between the pSWAP
operations used in the sections \ref{sec:spin} and \ref{sec:dd} above and the
one used in the operation of the quantum thermodynamic engine discussed here.
While the logical operation, a partial swap, is common to both versions, the
second implementation additionally features an energetic context. This energy
context is required in order to transform thermal energy into work and impacts
on the operation of this pSWAP. In particular, the logical operation can only
be executed when the implementation respects the conservation of physical
quantities, here, the conservation of energy. Furthermore, this conservation
of energy has to be achieved with a classical energy reservoir in order to
avoid entanglement between the wits and the reservoir. In fact, such an
entanglement would reduce the possible energy gain from the wit and generate
an additional waste heat, a feature we want to avoid with our implementation.

Two further remarks are in place at this point: i) No energy has been
transferred from the working cycle to the demon qubits. This is different from
the setups discussed in Refs.\ [\onlinecite{lloyd:97}] and
[\onlinecite{quan:06}], where part of the heat absorbed in the wits is wasted
by its transfer to the non-degenerate dits (with energy separation
$\Delta_\mathrm{d}$), leading to a reduction in efficiency $\eta =
1-\Delta_\mathrm{d}/\Delta_\mathrm{w}$ at best. The setup in Ref.\
[\onlinecite{delacruz:14}] makes use of energy-degenerate systems both for the
wits and dits and consumes purity of the dits in order to extract work from a
single reservoir.  Although, the latter setup is optimal (i.e., it extracts
the maximum possible work for a given purity), it involves a slow adiabatic
process. Specifically, in this machine, the heat is extracted in an isothermal
process during which the wit adiabatically absorbs energy from the reservoir
(i.e., this step involves a slow increase of $\Delta_\mathrm{w}$).  In
contrast, our engine operates at a constant $\Delta_\mathrm{w}$, i.e., the
isothermal process is replaced by an isochoric process in our cycle. The
advantage of the latter is that it does not require adiabatic heating, such
that the cycle time is restricted only by the time of the pSAWP operation,
assuming a fast equilibration time $T_1$ between the wit and the thermal
reservoir.  Assuming that the two CNOT operations can be done much faster than
the Rabi-period, the whole process of work extraction then requires only half
a Rabi-period of time. ii) The above energy balance ignores the energy needed
to purify the demon qubits for use in a next cycle.  This is justified by the
fact that no energy is transferred between the wits and dits, hence the engine
operates with mutually isolated energy- and entropy cycles.  The entropy
$\delta S$ that has been transferred to the dits can be dealt with in a
separate entropy cycle.  This additional step then accounts for the Second Law
of Thermodynamics (and Landauer's principle) which thus is fully respected by
the combined energy and entropy cycles.

In order to take the demon qubits back into the pure operational states
$|\!\!\uparrow\rangle$ and $|\!\!\downarrow\rangle$, one may bring them into
thermal contact with a second reservoir at a temperature $T_\mathrm{d}$ and
adiabatically disbalance their levels in order to establish a new pure state.
The energetic cost of disbalancing and rebalancing the levels is given by the
Helmholtz free energy $\delta F_\mathrm{d} = - T_\mathrm{d} \delta S$ of the
process. With $\delta S = S(\hat{r}_+) + S(\hat{r}_-) -S(\hat{r}_\uparrow) -
S(\hat{r}_\downarrow) = 2\ln Z + Q/T$, we have to invest an amount
$W_\mathrm{in} = 2 T_\mathrm{d} \ln Z +(T_\mathrm{d}/T)\, Q$ of work to
restore the dit's purities. We then arrive at a maximal overall efficiency
\begin{equation}\label{eq:eff}
   \eta_\mathrm{2-cy} = \frac{W_\mathrm{out} - W_\mathrm{in}}{Q}
   = 1-\frac{T_\mathrm{d}}{T}
   - \frac{T_\mathrm{d}}{\Delta_\mathrm{w}}\frac{\ln Z}{p_e},
\end{equation}
below the value $\eta_\mathrm{\scriptscriptstyle C} = 1 - T_\mathrm{d}/T$ of
the Carnot cycle, even for the idealized machine. The net work $W_\mathrm{out}
- W_\mathrm{in} = Q - T_\mathrm{d} \delta S < Q$ produced by the engine is
positive if $\beta_\mathrm{d} \Delta_\mathrm{w} > \beta \Delta_\mathrm{w} +
(1+e^{\beta\Delta_\mathrm{w}}) \ln(1+e^{-\beta\Delta_w})$.  Hence, a positive
work-yield requires that the temperature $T_\mathrm{d}$ of the demon reservoir
is lower than $T$. In fact, for a hot working reservoir
$\beta\Delta_\mathrm{w} \ll 1$, the minimal temperature of the cold reservoir
is defined by the energy-level spacing $\Delta_\mathrm{w}$, $\beta_\mathrm{d}
\Delta_\mathrm{w} > (2+ \beta\Delta_\mathrm{w}) \ln 2$ and only weakly depends
on $T$. The opposite regime with $\beta\Delta_\mathrm{w} \gg 1$ requires that
$\beta_\mathrm{d}\Delta_\mathrm{w}  > \beta\Delta_\mathrm{w} + 1$.

The operational separation into distinct energy and entropy cycles with no
energy transfer in between is the most interesting feature of this
quantum-thermodynamic engine. It implies that all the heat $Q$ absorbed from
the thermal reservoir by the working qubits in the energy cycle can be
extracted after the demon's action, $W_\mathrm{out}= Q$, and hence the energy
cylce locally runs with unit efficiency and does not produce any waste heat
(under ideal operation). All work reduction $W_\mathrm{in}$ enforced by the
Second Law is deferred to the entropy cycle which can be run in a separate
location or even at another time, e.g., by preparing a reservoir of demon
qubits which then can be `consumed' later in the operation of the engine.

In the above protocol, we have assumed that the two demon qubits are prepared
in perfectly pure states. If instead one allows a finite chaotic component in
$\hat{r}_\uparrow^\epsilon = \epsilon \mathds{1} + (1-2\epsilon)
|\!\!\uparrow\rangle \langle \uparrow\!\!|$ and $\hat{r}_\downarrow^\epsilon =
\epsilon \mathds{1} + (1-2\epsilon) |\!\!\downarrow\rangle \langle
\downarrow\!\!|$, then the resulting states of the wits take over this mixing
after the pSWAP operation, $\hat{\rho}_g^\epsilon = \epsilon \mathds{1} +
(1-2\epsilon) |g\rangle \langle g|$ and $\hat{\rho}_e^\epsilon = \epsilon
\mathds{1} + (1-2\epsilon) |e\rangle \langle e|$. As a result, the possible
gain in work is reduced: when extracting the energy from the second wit with
the help of a half-period Rabi pulse $\hat{u}_{\scriptscriptstyle 1/2}$, one
may excite rather than de-excite the wit, $\hat{u}_{\scriptscriptstyle 1/2}
|g\rangle = ie^{i\varphi} |e\rangle$, a process that occurs with probability
$\epsilon$. The average extracted work thus is reduced to $W_+^\epsilon =
(1-\epsilon) \Delta_\mathrm{w} - \epsilon \Delta_\mathrm{w} =
(1-2\epsilon)\Delta_\mathrm{w}$ and the total work gain is given by
$W_\mathrm{out}^\epsilon = 2\Delta_\mathrm{w} (p_e - \epsilon)$. On the other
hand, both wits remain excited with probability $\epsilon$ and hence the heat
absorbed in the subsequent cycle is reduced correspondingly, $Q^\epsilon =
2\Delta_\mathrm{w} (p_e-\epsilon)$, leading again to a maximal engine
efficiency $\eta^\epsilon = W_\mathrm{out}^\epsilon /Q^\epsilon = 1$.
Nevertheless, one can extract work out of the heat bath only when the dits are
initially more pure than the wits, i.e., for $\epsilon < p_e$.

When extending the discussion to include the entropy cycle, we have to
determine the work required to bring the dits back to their original mixed
states $\hat{r}_\uparrow^\epsilon$ and $\hat{r}_\downarrow^\epsilon$.  The
pSWAP operation in Eq.~(\ref{eq:wswap}) results in mixed dit states of the
form $\hat{r}_\pm^\epsilon = (1-\epsilon) \hat{r}_{\pm} + \epsilon
\hat{r}_{\mp}$. In order to restore the dits to their original states, one
needs to invest the external work $W_\mathrm{in}^\epsilon = T_\mathrm{d}
\delta S^\epsilon$, with  $\delta S^\epsilon = S(\hat{r}_+^\epsilon) +
S(\hat{r}_-^\epsilon) - S(\hat{r}_\uparrow^\epsilon) -
S(\hat{r}_\downarrow^\epsilon)$. Then, the net produced work
$W_\mathrm{2-cy}^\epsilon = W_\mathrm{out}^\epsilon - W_\mathrm{in}^\epsilon$
is given by
\begin{equation}\label{eq:power}
   W_\mathrm{2-cy}^\epsilon = 2\Delta_\mathrm{w}\biggl[p_e \!- \epsilon 
   - \frac{H[p_e \! + \epsilon(1\!-\!2p_e)] - H[\epsilon]}{\Delta_\mathrm{w} 
   \beta_\mathrm{d}} \biggr],
\end{equation}
where $H[x] = -x\ln x-(1-x)\ln(1-x)$ denotes the Shanon entropy.
Interestingly, $W_\mathrm{out}^\epsilon$ and $W_\mathrm{in}^\epsilon$ exhibit
different functional dependencies on the dit impurity $\epsilon$. E.g., for
$p_e \to 1/2$ (hot regime) the work $W_\mathrm{in}^\epsilon$ required for only
partial purification of the dit decreases more rapidly than the gain in
$W_\mathrm{out}^\epsilon$. Hence, one can find an optimal value
$\epsilon_\mathrm{\scriptscriptstyle W}$ which maximizes the work
(\ref{eq:power}) generated per cycle or, in other words, the engine power. On
the other hand, instead of maximizing the engine's power one can consider its
efficiency by comparing the net work $W_\mathrm{2-cy}^\epsilon$ with the
absorbed heat $Q^\epsilon = 2 \Delta_\mathrm{w} (p_e -\epsilon)$,
\begin{equation}\label{eq:eta}
   \eta_\mathrm{2-cy}^\epsilon = 1 - \frac{H[p_e+\epsilon(1-2p_e)]
   -H[\epsilon]}{\beta_\mathrm{d} \Delta_\mathrm{w} (p_e - \epsilon)},
\end{equation}
and find the optimal dit impurity $\epsilon_\eta$ maximizing the engine's
efficiency. In the following, we discuss the result of such an optimization 
and the emerging maximal power and efficiency of the optimized machine.

\begin{figure}[ht]
\begin{center}
\includegraphics[width=8truecm]{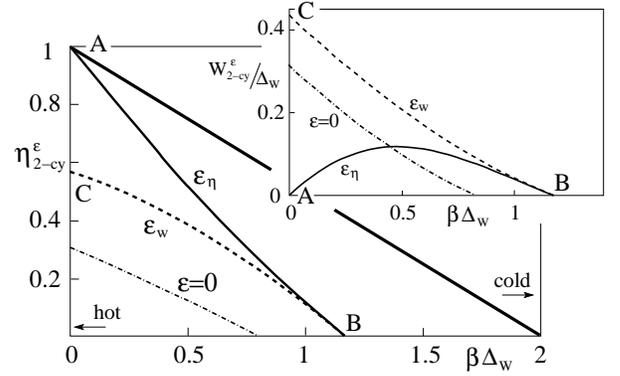}
\caption{Engine efficiencies $\eta_\mathrm{2-cy}^\epsilon$ and work per cycle
$W_\mathrm{2-cy}^\epsilon$ (inset) at a fixed demon temperature
$\beta_\mathrm{d} \Delta_\mathrm{w} = 2$ as a function of the working
temperature $\beta\Delta_\mathrm{w}$. Shown are the performances for the
idealized machine with $\epsilon = 0$ (dash-dotted line; this line decreases
with decreasing $\beta_\mathrm{d} \Delta_\mathrm{w}$ and disappears from the
plot at $\beta_\mathrm{d} \Delta_\mathrm{w} \leq 2\ln(2) \approx 1.39$) and
the efficiency- and power-optimized machines at $\epsilon = \epsilon_\eta$
(solid line) and $\epsilon = \epsilon_\mathrm{\scriptscriptstyle W}$
(dashed line).  The thick solid upper line describes the Carnot efficiency
$\eta_\mathrm{\scriptscriptstyle C} = 1 - T_\mathrm{d}/T$. Maximal efficiency
but vanishing power is attained in A. On approaching B, both efficiency and
power vanish. In C the power-optimized machine is characterized by both finite
efficiency and power.}
\label{fig:eff-work}
\end{center}
\end{figure}

The complete dependence of the optimal values
$\epsilon_\mathrm{\scriptscriptstyle W}$ and $\epsilon_\eta$ maximizing the
power and efficiency of the engine on the parameters $p_e$ and
$\beta_\mathrm{d} \Delta_\mathrm{w}$ has to be found numerically (see
appendix) and the result is shown in Fig.\ (\ref{fig:epsilon}).  For a hot
working reservoir with $\beta\Delta_w \ll 1$ and $p_e \to 1/2$ one can find
the approximate expressions
\begin{eqnarray}
   &&\epsilon_\mathrm{\scriptscriptstyle W} \approx 
   \bigl\{ 1 + \exp\bigl[ \beta_\mathrm{d} 
   \Delta_\mathrm{w} + H^\prime[p_e](1-2p_e)\bigr]\bigr\}^{-1},
   \label{eq:epspow}
   \\
   &&\epsilon_\eta \approx p_e - \frac12 (1-2p_e) + \frac23 (1-2p_e)^3,
   \label{eq:epseta}
\end{eqnarray}
where $H^\prime[x] = \ln[(1-x)/x]$ is the derivative of $H[x]$. Note that
$\epsilon_\eta$ depends only on $\beta$ (via $p_e$), while
$\epsilon_\mathrm{\scriptscriptstyle W}$ involves both temperatures $\beta$
and $\beta_\mathrm{d}$.  Quite surprisingly, in the hot regime with $p_e\to
1/2$, the optimal efficiency is reached at $\epsilon$-values close to $1/2$,
see also Eq.\ (\ref{eq:epseta}), i.e., for almost chaotic demon qubits.
However, at the same time the absorbed heat $Q^\epsilon$ goes to zero and so
does the work $W_\mathrm{2-cy}^\epsilon$, i.e., we deal with an optimal engine
but one that generates no power, see Fig.\ \ref{fig:eff-work}.  Indeed, the
optimal power is attained at lower values $\epsilon_\mathrm{\scriptscriptstyle
W} < 1/2$ and this value decreases further when the demon reservoir's
temperature $T_\mathrm{d}$ is lowered.

\begin{figure}[ht]
\begin{center}
\includegraphics[width=8truecm]{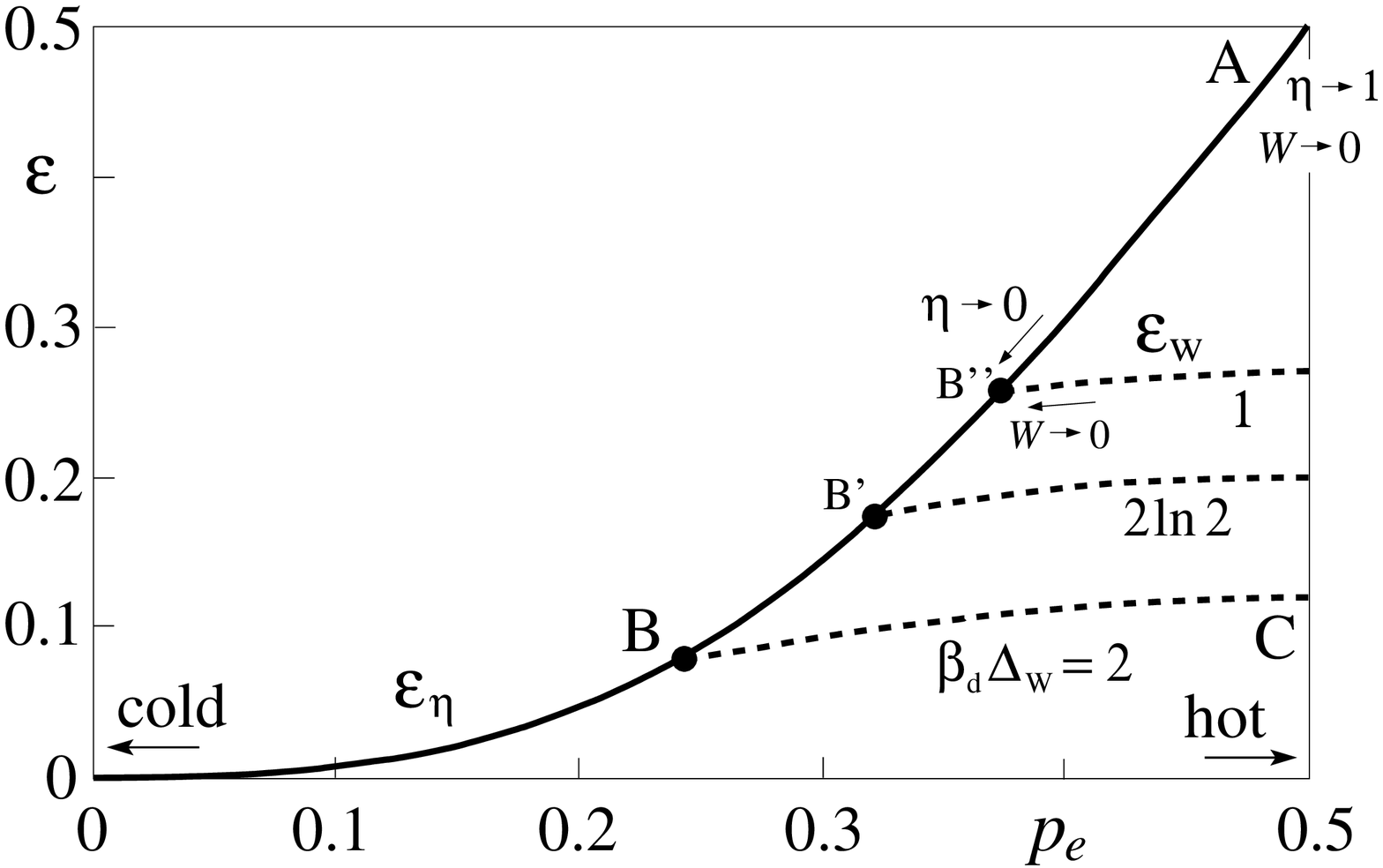}
\caption{Optimal values $\epsilon_\eta$ (solid line) and
$\epsilon_\mathrm{\scriptscriptstyle W}$ (dashed lines) for the dit impurity
$\epsilon$ versus occupation $p_e$ of excited states. The three dashed lines
correspond to different values $\beta_\mathrm{d} \Delta_\mathrm{w} = 1.0$
(upper curve), $\beta_\mathrm{d} \Delta_\mathrm{w} = 2\ln2$ below which the
ideal engine with $\epsilon = 0$ cannot produce positive work, and
$\beta_\mathrm{d} \Delta_\mathrm{w} =2$ (bottom curve).  At point A, the
efficiency $\eta_\mathrm{2-cy}^\epsilon$ approaches unity but the generated
work $W_\mathrm{2-cy}^\epsilon$ goes to zero. At the points point B, B' and
B'' both $\eta_\mathrm{2-cy}^\epsilon$ and $W_\mathrm{2-cy}^\epsilon$ approach
zero at fixed values of $\beta_\mathrm{d} \Delta_\mathrm{w}$ along the lines
$\epsilon_\eta(p_e)$ and $\epsilon_\mathrm{\scriptscriptstyle
W}(p_e,\beta_\mathrm{d})$ (note that $\epsilon_\eta$ does not depend on
$\beta_\mathrm{d}$ and $\eta$ approaches zero at different points along
$\epsilon_\eta(p_e)$ for different values of $\beta_\mathrm{d}
\Delta_\mathrm{w}$).  At point C, the work per cycle reaches a value
$W_\mathrm{2-cy}^\epsilon \approx 0.43\, \Delta_\mathrm{w}$ with an efficiency
$\eta_\mathrm{2-cy}^\epsilon = 0.57$. The points A, B, and C reappear in
Figure \ref{fig:eff-work} below.}
\label{fig:epsilon}
\end{center}
\end{figure}

The corresponding efficiencies $\eta_\mathrm{2-cy}$ and powers
$W_\mathrm{2-cy}^\epsilon$ for optimal dit impurities $\epsilon =
\epsilon_\mathrm{\scriptscriptstyle W}$ and $\epsilon = \epsilon_\eta$ are
shown in Fig.\ (\ref{fig:eff-work}) as a function of the inverse heating
temperature $\beta$ and for a demon reservoir with $\beta_\mathrm{d}
\Delta_\mathrm{w} = 2.0$. Also shown are the performances for an engine
operating with pure dit states $\epsilon = 0$, which is always
underperforming with respect to the efficiency.  Moreover, for a hot demon
reservoir with $\beta_\mathrm{d} \Delta_\mathrm{w} \leq 2\ln(2) \approx 1.39$,
the ideal engine with $\epsilon = 0$ cannot produce positive work. On the
other hand, the optimal engines with either $\epsilon = \epsilon_\eta$ or
$\epsilon = \epsilon_\mathrm{\scriptscriptstyle W}$ can yield positive work
for any temperature of the demon reservoir, though the operating regime
$\beta\in[0,\beta_\mathrm{d}/2]$ is small at high temperatures
$\beta_\mathrm{d}\Delta_\mathrm{w} \to 0$, see appendix. In the opposite
situation of a cold demon reservoir $\beta_\mathrm{d} \Delta_\mathrm{w} \gg 1$
the temperature of the working reservoir is bound by $\beta \Delta_\mathrm{w}
< \beta_\mathrm{d} \Delta_\mathrm{w} - 1$. The three efficiency curves
$\eta_\mathrm{2-cy}^\epsilon(\beta)$ for $\epsilon =0,~\epsilon_\eta,~
\epsilon_\mathrm{\scriptscriptstyle W}$ then approach one another near the
critical value $\beta \Delta_\mathrm{w} \approx \beta_\mathrm{d}
\Delta_\mathrm{w} - 1$. Finally, as is obvious from Fig.\
(\ref{fig:eff-work}), the $\epsilon_\eta$ and
$\epsilon_\mathrm{\scriptscriptstyle W}$ optimized engines never attain the
Carnot efficiency, except for a very hot working reservoir, where
\begin{equation}
  \eta_\mathrm{2-cy}^\epsilon(\beta \to 0,\beta_\mathrm{d}) \approx 1
   - 2 \frac{T_\mathrm{d}}{T} \Bigl( 1 - \frac{(\beta\Delta_\mathrm{w})^2}6
   \Bigr),
\end{equation}
at $\epsilon = \epsilon_\eta$. In order to reach the Carnot efficiency, one has to
use working qubits with a tunable level spacing and replace our isochoric heating
by an adiabatic isothermal process as in Ref.\ [\onlinecite{delacruz:14}].

\section{Summary and conclusion} \label{sec:sc}

Quantum mechanics allows for processes that are classically forbidden. This
idea can be pursued in the context of thermodynamics and the Second Law. In
this work, we have investigated the possibility for an energy-isolated system
to undergo an evolution with a decreasing entropy. In order to render this
task non-trivial, one has to replace the classical concept of an `isolated
system' with an energy-isolated but open quantum system that can entangle with
its (micro-)environment.  This implies allowing a phase-exchange with the
environment.  Furthermore, the environment and its interaction with the system
has to be properly engineered---rather than an uncontrolled macroscopic
environment inducing incoherence, we had to consider a properly designed
micro-environment.

With these prerequisites in mind and starting from the
quantum-information-theoretic insight that unital quantum channels produce an
evolution with non-decreasing entropy, we have searched for an
energy-conserving non-unital quantum channel to reach our goal. Furthermore,
we have considered a minimal two-qubit mesoscopic setting with a scattering
electron (a flying system-qubit) that interacts with a micro-environment in
the form of a spin (the demon qubit), interacting with the electron's current
via its magnetic field.  This approach naturally has led us to a process where
the electron--spin (or system--demon) interaction generates two non-commuting
rotations of the spin.  An alternative setup with the spin replaced by a
double-dot qubit allows for two equal interaction events but requires a qubit
rotation in between, ultimately producing two non-commuting operations on the
double-dot as well.

The analysis of these specific mesoscopic examples in terms of a quantum
circuit has revealed a general scheme or algorithm that exerts a partial SWAP
operation between the system and demon qubits.  Swapping the system state
against a purer or less mixed environmental state naturally explains the
system's reduction in entropy or increase in coherence, where the latter is
more easily observed in a direct interference experiment.  Starting and ending
with product states, the pSWAP process ultimately does not entangle the system
with its environment. While the system state can be chosen arbitrary, i.e., a
pure superposition state or any mixed state, the micro-environment has to be
properly prepared, implying the provision of information to the system.
Furthermore, within our scheme, the requirement of no net energy exchange
between the system and the micro-environment is guaranteed by using an
environmental qubit with energy-degenerate states. Note that the conservation
of the subsystem's individual energies is a subtle issue as the partition of
the interaction energy is not clear. We thus have used either a phase
interaction, where this question does not appear at all, or have focused on
the energy conservation between initial and final states.

The insight that the system's entropy decrease is based on a SWAP of states
makes sure that nothing mysterious happens with the Second Law of
thermodynamics, as the overall entropy of system plus micro-environment is
conserved. Rather, the micro-environment assumes the role of a quantum Maxwell
demon and in this terminology the honoring of the Second Law corresponds to
Landauer's principle applied in the restauration of the demon qubits.

A number of designs for quantum thermodynamic engines have been proposed,
including such that make use of SWAP operations. Having focused on
energy-isolated systems, our approach naturally has lead us to a design where
the working (or system) cycle is energy-separated from the demon cycle. This
feature provides the possibility to feed a heat engine with one local thermal
reservoir, while providing purity from a distant source that can be handled
independently. Such a design is useful when the thermal- to directed-energy
conversion should be free from (uncontrolled) waste heat. As an alternative,
one might trade the entropy in the thermally excited state against entropy
stored in other, e.g., orbital \cite{scully:01}, degrees of freedom of the
same system.  Analyzing the practical aspects of our design, we have found
that the best machine is not the ideal one involving pure demon qubits, but
that some degree of incoherence is favorable, both in terms of efficiency and
power. The ideal machine with a vanishing demon impurity $\epsilon$ is
recovered when optimizing the engine for maximal work at small demon
temperatures $T_\mathrm{d}$.

Over many decades, Maxwell's demon has been a fascinating concept, despite his
imprisonment by Landauer's principle.  Our quantum Maxwell demon based on
degenerate demon qubits and allowing for an energy-isolated operation of the
working cycle is possibly as close as one could imagine to Maxwell's original
idea of a demon.

\begin{acknowledgements}

We thank Renato Renner for discussions and acknowledge financial support from
the National Science Foundation through the National Center of
Competence in Research on Quantum Science and Technology (QSIT), the Pauli
Center for Theoretical Studies at ETH Zurich, and the RFBR Grant No.\
14-02-01287.

\end{acknowledgements}

\appendix


\section{Quantum engine optimization}\label{app:engine}

The optimal $\epsilon$-values for the demon qubits which maximize either the
engine work or power ($\epsilon = \epsilon_\mathrm{\scriptscriptstyle W}$) or
engine efficiency ($\epsilon = \epsilon_\eta$) can be found by differentiating
the corresponding target quantities ($W_\mathrm{2-cy}^\epsilon$ or
$\eta_\mathrm{2-cy}^\epsilon$) with respect to $\epsilon$, see
Eqs.\ (\ref{eq:power}) and~(\ref{eq:eta}), with the relations
\begin{equation}
  (1-2p_e) H^\prime[p_e + \epsilon (1-2p_e)] - H^\prime[\epsilon] 
   = -\beta_\mathrm{d}
   \label{eq:powcon}
\end{equation}
providing $\epsilon_\mathrm{\scriptscriptstyle W}$ and
\begin{eqnarray}
   &&\bigl[(1-2p_e) H^\prime[p_e + \epsilon (1-2p_e)] 
   - H^\prime[\epsilon]\bigr](p_e -\epsilon)
   \nonumber\\
   &&\qquad\quad + H[p_e + \epsilon (1-2p_e)] - H[\epsilon] = 0
   \label{eq:effcon}
\end{eqnarray}
the value for $\epsilon_\eta$.  These equations can be solved in the
high-temperature regime $\beta\Delta_\mathrm{w} \gg 1$ where $p_e
\to 1/2$ and which is governed by the small parameter 
\begin{equation}
   \xi = 1 - 2p_e = \frac{1 - e^{-\beta\Delta_\mathrm{w}}}
   {1+e^{-\beta\Delta_\mathrm{w}}}.
\end{equation}
Assuming $H^\prime[p_e + \epsilon (1-2p_e)] = H^\prime[p_e + \epsilon\xi]
\approx H^\prime[p_e]$ in (\ref{eq:powcon}), one straightforwardly arrives at
the result for $\epsilon_\mathrm{\scriptscriptstyle W}$ as given by
Eq.\ (\ref{eq:epspow}).

Next, we find $\epsilon_\eta$ from Eq.\ (\ref{eq:effcon}) using the ansatz
$\epsilon_\eta = p_e - \delta p = (1-\xi)/2 - \delta p$ with $\delta p > 0$,
\begin{align}
  &\delta p \, \Bigl[\xi H^\prime\bigl[(1\!-\! \xi^2)/2-\xi \delta p \bigr] 
                        - H^\prime\bigl[(1\! -\!\xi)/2 -\delta p \bigr]\Bigr]
   \nonumber\\
   &+ H\bigl[(1\!-\!\xi^2)/2-\xi \delta p \bigr] 
           - H\bigl[(1\!-\!\xi)/2 -\delta p \bigr] = 0.
\end{align}
Expanding $H[x]$ and $H^\prime[x]$ near $x = 1/2$ (where $H^\prime[1/2] = 0$)
up to the fourth and third order, respectively, one arrives at the algebraic
equation
\begin{equation}
   48 (\delta p)^4 + 64 \xi (\delta p)^3 + 24(\delta p)^2 - 6\xi^2 
   + 5\xi^4 = 0,
\end{equation}
with the solution $\delta p \approx \xi/2 - 2\xi^3/3 + O(\xi^5)$ producing the
result Eq.\ (\ref{eq:epseta}).

Finally, we find the minimal temperature $T_\mathrm{m}$ (or maximal
$\beta_\mathrm{m}$) where the engine provides positive work. For a fixed demon
parameter $\beta_\mathrm{d} \Delta_\mathrm{w}$, the corresponding constraint
on $p_e$ and $\epsilon$ has the form (see Eq.\ (\ref{eq:power})),
\begin{equation}
   H[p_e + \epsilon (1-2p_e)] - H[\epsilon] = \beta_\mathrm{d}\Delta_\mathrm{w}
   (p_e - \epsilon).
   \label{eq:poscon}
\end{equation}
For a power-optimized engine, $\epsilon = \epsilon_\mathrm{\scriptscriptstyle
W}$, and combining (\ref{eq:poscon}) with (\ref{eq:powcon}) one immediately
arrives at Eq.\ (\ref{eq:effcon}). This implies that near $\beta_\mathrm{m}$
both curves $\epsilon_\mathrm{\scriptscriptstyle W}$ and $\epsilon_\eta$
coincide, resulting in the same engine efficiencies near the critical
temperature. This observation explains the merging curves in Fig.\
(\ref{fig:eff-work}) near $\beta_\mathrm{m}$.  In the high-temperature regime
$\beta_\mathrm{d} \Delta_\mathrm{w}, \beta\Delta_\mathrm{w} \ll 1$ one can
substitute $\epsilon_\eta$ as given by Eq.\ (\ref{eq:epseta}) into
(\ref{eq:poscon}); an expansion with respect to $\xi \ll 1$ provides the
approximate solution $\beta_\mathrm{m} \approx \beta_\mathrm{d} / 2$.

\end{document}